\documentclass[12pt,preprint]{emulateapj}

\slugcomment{Astrophysical Journal}

\shorttitle{Buried AGNs in a complete sample of nearby ULIRGs}
\shortauthors{Imanishi}

\begin{document}


\title{Spitzer Infrared Low-Resolution Spectroscopic Study of Buried
AGNs in a Complete Sample of Nearby Ultraluminous Infrared Galaxies}    


\author{Masatoshi Imanishi\altaffilmark{1}}
\affil{National Astronomical Observatory, 2-21-1, Osawa, Mitaka, Tokyo
181-8588, Japan}
\email{masa.imanishi@nao.ac.jp}

\author{Roberto Maiolino}
\affil{INAF - Osservatorio Astronomico di Roma,
via di Frascati 33, I-00040 Monte Porzio Catone, Roma, Italy}

\and

\author{Takao Nakagawa}
\affil{Institute of Space and Astronautical Science, Japan Aerospace
Exploration Agency, 3-1-1 Yoshinodai, Sagamihara, Kanagawa 229-8510,
Japan}

\altaffiltext{1}{Department of Astronomy, School of Science, Graduate
University for Advanced Studies, Mitaka, Tokyo 181-8588}

\begin{abstract}
We present the results of {\it Spitzer} IRS low-resolution infrared
5--35 $\mu$m spectroscopy of 17 nearby ULIRGs at $z <$ 0.2, optically
classified as non-Seyferts. 
The presence of optically elusive, but intrinsically luminous, buried
AGNs is investigated, based on the strengths of polycyclic aromatic
hydrocarbon emission and silicate dust absorption features detected in
the spectra. 
The signatures of luminous buried AGNs, whose intrinsic luminosities
range up to $\sim$10$^{12}$L$_{\odot}$, are found in eight sources. 
We combine these results with those of our previous research to
investigate the energy function of buried AGNs in a complete sample of
optically non-Seyfert ULIRGs in the local universe at $z <$ 0.3 (85
sources).  
We confirm a trend that we previously discovered: that buried AGNs are
more common in galaxies with higher infrared luminosities.
Because optical Seyferts also show a similar trend, we 
argue more generally that
the energetic importance of AGNs is intrinsically higher in more
luminous galaxies, suggesting that the AGN-starburst connections are 
luminosity-dependent. 
This may be related to the stronger AGN feedback scenario 
in currently more massive galaxy systems, as a
possible origin of the galaxy downsizing phenomenon. 
\end{abstract}


\section{Introduction}

Ultraluminous infrared galaxies (ULIRGs) are characterized by
a spectral energy distribution dominated by infrared emission, 
and an absolute infrared luminosity with 
L$_{\rm IR}$ $>$ 10$^{12}$L$_{\odot}$ \citep{san88a,sam96}.  
This means that in ULIRGs, (1) very luminous energy sources with 
L $>$ 10$^{12}$L$_{\odot}$ are present, 
(2) the energy sources are hidden by dust, which absorbs most of the 
primary radiation, and 
(3) the heated dust grains then radiate this energy as infrared dust
emission.   
The dust-obscured energy sources are nuclear fusion occurring
inside rapidly formed stars (starbursts) and/or the release of gravitational
energy generated by spatially compact, mass-accreting, supermassive black
holes (SMBHs) (i.e., AGN activity).  
Because the ULIRG population becomes very important at $z >$ 1, in terms
of cosmic infrared radiation density \citep{cap07}, understanding the
hidden energy sources of ULIRGs is closely coupled with clarifying the
connection between star formation and SMBH growth in the dust-obscured
galaxy population of the early universe.  
Because distant ULIRGs are generally faint, detailed studies of nearby 
($z <$ 0.3) ULIRGs continue to play an important role in understanding
the nature of the ULIRG population of the universe.

Nearby ULIRGs have two important observational properties. 
First, the fraction of optical Seyferts (optically identified AGNs) 
is known to be substantially greater in ULIRGs, compared with galaxies
with lower infrared luminosities (L$_{\rm IR}$ $<$ 
10$^{12}$L$_{\odot}$) \citep{vei99,got05}.
Second, although galaxies with L$_{\rm IR}$ $<$ 10$^{12}$L$_{\odot}$
generally exhibit strong, spatially extended infrared emission
originating from stars distributed inside them, the infrared dust
emission of nearby ULIRGs is dominated by a spatially compact feature
\citep{soi00,soi01}, suggesting that a large amount of dust
is concentrated in the nuclear regions of ULIRGs, and 
heated by spatially compact energy sources.

Because the nuclear dust concentration increases in ULIRGs
\citep{sam96,soi00}, the putative AGNs in ULIRG's cores are 
surrounded by a large amount of gas and dust, and 
ionizing UV radiation from the AGN can be blocked at the inner part 
($<$10 pc) in virtually all lines-of-sight \citep{hop05,idm06,ima07a}. 
Such {\it buried} AGNs lack well-developed narrow-line-regions (NLRs) at
10--1000 pc scale, the primary sources of forbidden emission lines, 
and so are elusive through the conventional optical and infrared 
spectroscopic classification, looking for high-excitation forbidden
emission lines from the NLRs \citep{vei87,kew06,arm07,far07}.
Low-resolution infrared 2.5--35 $\mu$m spectroscopy is an effective tool
for studying such elusive \citep{mai03} buried AGNs, for the following
reasons. 
First, the effects of dust extinction are sufficiently small
($<$0.06A$_{\rm V}$ Nishiyama et al. 2008; 2009) that buried AGNs are
detectable.   
Second, in a normal starburst with moderate metallicity ($>$0.3 solar),
consisting of UV-emitting HII regions, molecular gas and dust, and
photo-dissociation regions (PDRs), the PAHs in the PDRs are excited by
far-UV photons from stars without being destroyed, so strong PAH
emission is usually detected \citep{sel81,wu06}.
However, PAHs in close proximity to an AGN are destroyed by the 
high-energy X-ray radiation of the AGN \citep{voi92}, so that PAH
emission is virtually absent from a pure AGN, as demonstrated from 
observations \citep{moo86,roc91,gen98,imd00}.
In a galaxy hosting an AGN, strong PAH emission can be observed, if 
PAHs survive in the regions shielded from the high-energy radiation of
the AGN, and PAH-exciting far-UV photons from local energy sources
(i.e., stars) are available there.
Finally, in a normal starburst, stellar energy sources and gas/dust are
spatially well mixed \citep{pux91,mcl93,for01}, so that the absolute
optical depths of dust absorption features in the infrared spectra cannot
exceed certain thresholds, whereas the optical depths can be arbitrarily
large in a buried AGN, because the energy source (a compact
mass-accreting SMBH) is more centrally concentrated than
dust/gas \citep{im03,idm06,ima07a}.
Thus, in principle, starbursts and buried AGNs are distinguishable
on the basis of the strengths of PAH emission and dust absorption
features in the infrared spectra.

From the low-resolution infrared 2.5--35 $\mu$m spectra of a large
number of nearby ULIRGs, obtained with {\it ISO}, {\it Spitzer} and 
{\it AKARI} infrared satellites, the detectable buried AGN fraction has
indeed been found to increase in ULIRGs, compared with galaxies with
lower infrared luminosities \citep{tra01,ima08,ima09,vei09,nar09}.
However, because the sample is not statistically complete, some bias and 
ambiguity could remain.
Observations of a complete sample will clearly lead to a better
understanding of the true nature of the nearby ULIRG population.

In this paper, we present the {\it Spitzer} IRS low-resolution infrared
5--35 $\mu$m spectra of previously unobserved, optically non-Seyfert
ULIRGs at $z <$ 0.3.
By combining these data with our previous results \citep{ima07a,ima09},
we can carry out a systematic investigation of the energetic importance
of buried AGNs in nearby ULIRGs, based on the complete sample. 
Throughout this paper, H$_{0}$ $=$ 75 km s$^{-1}$ Mpc$^{-1}$,
$\Omega_{\rm M}$ = 0.3, and $\Omega_{\rm \Lambda}$ = 0.7 are adopted
for the sake of consistency with our previously published papers.

\section{Targets}

The ULIRGs in the {\it IRAS} 1 Jy sample \citep{kim98} are our targets.
This sample consists of 118 ULIRGs at $z <$ 0.3, whose {\it IRAS} 60
$\mu$m fluxes are larger than 1 Jy. Of these, 33 are
optically classified as Seyferts, and 85 as non-Seyferts
\citep{vei99}.
The investigation of optically elusive, but intrinsically luminous,
buried AGNs in the 85 optically non-Seyfert ULIRGs is our main objective. 
The 85 sources consist of 43 optically LINER, 32 optically HII-region, 
and 10 optically unclassified ULIRGs \citep{vei99}.

\citet{ima07a} investigated the buried AGN fraction in the complete
sample of 48 ULIRGs at $z <$ 0.15, classified optically as LINERs and 
HII-regions, in the IRAS 1 Jy sample, on the basis of {\it Spitzer}
low-resolution infrared 5--35 $\mu$m spectra.  
These authors found strong signatures of luminous buried AGNs in a
sizable fraction (16/48; 33\%) of the observed ULIRGs.
Four optically unclassified ULIRGs at $z <$ 0.15 were not studied. 

\citet{ima09} extended the {\it Spitzer} IRS low-resolution infrared
spectroscopic study to ULIRGs at $z >$ 0.15, classified optically as
non-Seyferts (LINERs, HII-regions, and unclassified), in the IRAS 
1 Jy sample, by analyzing 
the spectra of 20 sources available at that time.
As a result of this extension, a large number of ULIRGs with L$_{\rm
IR}$ $\geq$ 
10$^{12.3}$L$_{\odot}$ are now included, so that a meaningful
investigation of the buried AGN fraction as a function of galaxy
infrared luminosity has become possible, by dividing galaxies into
different infrared luminosity classes.  
Specifically, the classes are defined by L$_{\rm IR}$ $<$
10$^{12}$L$_{\odot}$,  
10$^{12}$L$_{\odot}$ $\leq$ L$_{\rm IR}$ $<$ 10$^{12.3}$L$_{\odot}$, 
and L$_{\rm IR}$ $\geq$ 10$^{12.3}$L$_{\odot}$.   
\citet{ima09} found a trend that the detectable buried AGN
fraction systematically increases with increasing galaxy infrared
luminosity.  
However, the ULIRG sample was not statistically complete, because 
unobserved, optically non-Seyfert ULIRGs remained in the IRAS 1 Jy
sample.   

Among 33 ULIRGs at $z >$ 0.15 optically classified as non-Seyferts 
in the IRAS 1 Jy sample, 
five LINER, six HII-region, and two optically unclassified ULIRGs were
not studied in \citet{ima09}.
Additionally, four optically unclassified ULIRGs at $z <$ 0.15 were  
not investigated by \citet{ima07a} and \citet{ima09}.
In total, there were 17 (20\% of the total 85 sources) optically
non-Seyfert ULIRGs without published {\it Spitzer} IRS 5--35 $\mu$m
low-resolution spectra in the {\it IRAS} 1 Jy sample.
Because {\it Spitzer} IRS 5--35 $\mu$m low-resolution spectroscopy of
one source (IRAS 08474+1813) was scheduled in Cycle 3, we observed the
remaining 16 ULIRGs on our own, during Spitzer Cycle 5. 
Table 1 summarizes the properties of these 17 ULIRGs.

\section{Observations and Data Analysis}

Observations of all 17 ULIRGs were performed with the Infrared
Spectrograph (IRS) \citep{hou04} on board the Spitzer Space Telescope
\citep{wer04}.  All four modules, Short-Low 2 (SL2; 5.2--7.7 $\mu$m)
and 1 (SL1; 7.4--14.5 $\mu$m), and Long-Low 2 (LL2; 14.0--21.3 $\mu$m)
and 1 (LL1; 19.5--38.0 $\mu$m) were used to obtain the full 5--35 $\mu$m
low-resolution (R $\sim$ 100) spectra.  
Table 2 contains details of the observation log.
The slit width was 3$\farcs$6 or two pixels for SL2 (1$\farcs$8
pixel$^{-1}$) and 3$\farcs$7 or $\sim$2 pixels for SL1 (1$\farcs$8
pixel$^{-1}$).  For LL2 and LL1, the slit widths were 10$\farcs$5 and
10$\farcs$7, respectively, corresponding to $\sim$2 pixels for both
LL2 (5$\farcs$1 pixel$^{-1}$) and LL1 (5$\farcs$1 pixel$^{-1}$).  

The latest pipeline-processed data products available at the time of our
analysis were used.  Frames taken at position A were subtracted from
those taken at position B to remove background emission, consisting mostly 
of zodiacal light. The spectra were then extracted in the standard manner. 
Apertures with 4--6 pixels were employed for SL and LL data.
All sources were dominated by spatially compact emission whose size 
is similar to the PSF of SL and LL.
The spectra extracted for the A and B positions were then summed.
Wavelength and flux calibrations were made on the basis of 
the {\it Spitzer} pipeline-processed data.
For the SL1 spectra, the data at $\lambda_{\rm obs}$ $>$ 14.5 $\mu$m in 
the observed frame are invalid (Infrared Spectrograph Data Handbook,
version 1.0), and thus they were removed.  For the LL1 spectra, the data
at $\lambda_{\rm obs}$ $>$ 35 $\mu$m were not used, because they are  
noisy and not necessary for our scientific discussions.

For the flux calibration, we did not re-calibrate our spectra
using the {\it IRAS} measurements at 12 $\mu$m and 25 $\mu$m, because 
the {\it IRAS} 12 $\mu$m and/or 25 $\mu$m fluxes are only upper limits 
in many sources (Table 1).
Hence, the absolute flux calibration is dependent on the accuracy of the
pipeline-processed data, which is taken to be better than 20\% 
for SL and LL (Infrared Spectrograph Data Handbook). This level of
flux uncertainty does not significantly affect our main conclusions. 
For ULIRGs with {\it IRAS} 25 $\mu$m detection, we confirmed that the
{\it Spitzer} IRS 25 $\mu$m flux agrees with the {\it IRAS} 25
$\mu$m data within 30\%. 
For ULIRGs with {\it IRAS} 25 $\mu$m non-detection, the measured
{\it Spizter} IRS 25 $\mu$m flux is always smaller than the 
{\it IRAS} 25 $\mu$m upper limits.

For all ULIRGs, flux discrepancies between SL1 and LL2 were discernible,
ranging from 40\% to a factor of $\sim$2, and the SL1 flux
(3$\farcs$7 wide slit) was always smaller than the LL2 flux
(10$\farcs$5). 
The discrepancies were generally large in double-nuclei ULIRGs with
separations of a few to several arcsec (marked sources in the column 1
of Table 2), possibly because the LL2 spectra 
cover the emission from both nuclei and a diffuse component, while the
SL1 spectra only probe the emission from a brighter nucleus. 
We adjusted the SL1 (and SL2) flux to match the LL2 flux, to minimize 
missing PAH fluxes.
Appropriate spectral binning with 2, 4, or 8 pixels was applied to
reduce the scattering of data points particularly at SL2 (5.2--7.7
$\mu$m) and LL2 (14.0--21.3 $\mu$m) for faint ULIRGs, and at
$\lambda_{\rm obs}$ $\sim$ 10 $\mu$m for ULIRGs that show deep 9.7
$\mu$m silicate dust absorption features.

\section{Results}

The infrared 5--35 $\mu$m low-resolution spectra of the
17 observed ULIRGs are shown in Figure 1. 
The full 5--35 $\mu$m spectra of these ULIRGs are published here for the
first time. 

The spectra in Figure 1 are useful for observing the overall 5--35 $\mu$m
spectral shapes and broad 9.7 $\mu$m and 18 $\mu$m silicate dust
absorption features, but not for observing PAH emission.  
Following \citet{ima07a} and \citet{ima09}, Figure 2 presents enlarged
spectra at $\lambda_{\rm obs}$ = 5.2--14.5 $\mu$m to better exhibit the
PAH emission properties.

\subsection{PAH emission}

All ULIRGs in Figure 2 show clear signs of PAH emission at
$\lambda_{\rm rest}$ = 6.2 $\mu$m, 7.7 $\mu$m, and 11.3 $\mu$m 
in the rest frame. To
estimate the strengths of these PAH emission, we adopted a
linear continuum, following \citet{ima07a} and \citet{ima09}.
Data points at slightly shorter and longer wavelengths than the 6.2 $\mu$m,
7.7 $\mu$m, and 11.3 $\mu$m PAH emission were used to
determine linear continuum levels, which are shown as solid lines in
Figure 2. 
PAH emission features above the adopted continuum levels were fitted
with Gaussian profiles. 
The estimated rest-frame equivalent widths (EW$_{\rm PAH}$) and
luminosities of the 6.2 $\mu$m, 7.7 $\mu$m, and 11.3 $\mu$m PAH 
emission are summarized in Table 3.

As noted by \citet{ima07a} and \citet{ima09}, we estimated the strength
of the 7.7 $\mu$m PAH emission in such a way that the
uncertainty caused by the strong, broad 9.7 $\mu$m silicate dust
absorption feature is minimized. 
This definition of the 7.7 $\mu$m PAH emission strength is
significantly different from that used in previous papers (e.g.,
Genzel et al. 1998).
We will not use 7.7 $\mu$m PAH emission strengths in our main
discussions, because of the difficulty of comparing our data with data
in the literature, as well as possible large uncertainties in the 7.7
$\mu$m PAH strengths.

\subsection{Silicate absorption}

To estimate the strengths of the silicate dust absorption features, 
we use $\tau_{9.7}'$ and $\tau_{18}'$, defined by \citet{ima07a}.
The value of $\tau_{9.7}'$ is the optical depth of the 9.7 $\mu$m silicate
absorption feature, relative to a power-law continuum, determined from data
points at $\lambda_{\rm rest}$ = 7.1 $\mu$m and 14.2 $\mu$m, to minimize
the effects of PAH emission.  
The value of $\tau_{18}'$ is the optical depth of the 18 $\mu$m silicate
absorption feature, relative to a power-law continuum determined from
data points at $\lambda_{\rm rest}$ = 14.2 $\mu$m and 24 $\mu$m.  
These continua are shown as dotted lines in Figure 1. 
Because these continuum levels are determined using data points just
outside the 9.7 $\mu$m and 18 $\mu$m features, and close to the
absorption peaks, the measured optical depths are taken as true dips
caused by the silicate dust absorption feature.  
The $\tau_{9.7}'$ and $\tau_{18}'$ values were estimated from several data
points, close to the absorption peaks. 
The $\tau_{9.7}'$ values for all the ULIRGs are shown in Table 4 (column 2).
The $\tau_{18}'$ values are also shown in Table 4 (column 3) for ULIRGs
with clearly detected 18 $\mu$m silicate absorption features.

\subsection{Ice absorption}

Many ULIRGs display dips on the shorter wavelength side of the 6.2
$\mu$m PAH emission, which we ascribe to the 6.0 $\mu$m H$_{2}$O ice
absorption feature (bending mode). 
Figure 3 presents enlarged spectra at $\lambda_{\rm obs}$ = 5.2--9 $\mu$m
for ULIRGs with clearly detected ice absorption features. 
The spectrum of IRAS 14202+2615 is also included, as an example of 
non-detection. 
The observed optical depths ($\tau_{6.0}$) are summarized in Table 5 
for detected sources.
The detection rate of this 6.0 $\mu$m ice absorption feature is
substantially higher in optically LINER ULIRGs (5/5; 100\%) than in
HII-region ULIRGs (1/6; 17\%).

\section{Discussion}

For the sake of consistency, we use the criteria employed by \citet{ima07a}
and \citet{ima09} to study buried AGNs in optically non-Seyfert ULIRGs. 
We first search for ULIRGs with luminous buried AGN signatures 
in infrared 5--35 $\mu$m spectra, and then
estimate the extinction-corrected intrinsic luminosities of buried 
AGNs, based on the observed fluxes at $\sim$10 $\mu$m and dust
extinction toward the 10 $\mu$m continuum emitting regions 
\citep{ima07a,ima09}.

Several other methods of evaluating the energetic importance of buried
AGNs, based on {\it Spitzer} IRS infrared low-resolution spectra, have
been proposed in the literature \citep{vei09,nar09}.    
In the methods of \citet{vei09}, AGN and starburst zero-points were 
derived from {\it unobscured} AGNs and starburst galaxies,
respectively.
Then, relative buried AGN contributions to the {\it observed} fluxes of 
continuum and line emission at infrared 5--35 $\mu$m were derived. 
No dust extinction correction was applied to estimate buried AGN 
luminosities. 
Even though dust extinction at 5--35 $\mu$m is smaller than that 
at shorter wavelengths, using the {\it observed} fluxes could 
underestimate the intrinsic buried AGN luminosities, because emission
from buried AGNs is more highly flux-attenuated than the surrounding
starbursts in individual ULIRGs, as well as unobscured AGNs used to
determine the AGN zero-points. 

\citet{nar09} applied dust extinction correction to estimate the 
intrinsic buried AGN luminosities, but their modeling utilized only
5--8 $\mu$m spectra.   
Our methods use a wider-wavelength range (5--35 $\mu$m) of {\it Spitzer}
IRS spectra, and dust extinction is also taken into account to estimate
the buried AGN luminosities.  

\subsection{Detected modestly obscured starbursts}

Aside from the strong 9.7 $\mu$m silicate absorption peak, the flux
attenuation of the continuum emission at $\lambda_{\rm rest}$ $>$ 5
$\mu$m is small ($<$0.5 mag) for dust extinction with A$_{\rm V}$ $<$ 20 
mag \citep{nis08,nis09}.
Thus, the observed PAH emission luminosities at $\lambda_{\rm rest}$ $>$
5 $\mu$m can be used to roughly estimate the intrinsic luminosities of
modestly obscured (A$_{\rm V}$ $<$ 20 mag) PAH-emitting normal
starbursts (with PDRs and modest metallicity), provided that the PAH
emission and infrared luminosities proportionally correlate in the
starbursts \citep{pee04,soi02}.  
Since the metallicity of ULIRGs is estimated to be solar or even higher 
\citep{rup08,vei09}, the assumption of PAH-emitting starbursts is 
valid in ULIRGs \citep{wu06,mad06,oha06,smi07}. 
Table 3 (columns 8 and 9) lists the values of the 6.2 $\mu$m PAH to
infrared luminosity ratio, L$_{\rm 6.2PAH}$/L$_{\rm IR}$, and the 11.3
$\mu$m PAH to infrared luminosity ratio, L$_{\rm 11.3PAH}$/L$_{\rm IR}$. 
In normal starburst galaxies with modest dust obscuration
(A$_{\rm V}$ $<$ 20 mag), the ratios are estimated to be 
L$_{\rm 6.2PAH}$/L$_{\rm IR}$ $\sim$ 3.4 $\times$ 10$^{-3}$ \citep{pee04}
and L$_{\rm 11.3PAH}$/L$_{\rm IR}$ $\sim$ 1.4 $\times$ 10$^{-3}$
\citep{soi02} 
\footnote{
\citet{smi07} derived high PAH to infrared luminosity ratios, by
including underlying plateau components as PAH fluxes. 
We employ the ratios obtained by \citet{pee04} and \citet{soi02},
because of their similar continuum choices to ours.  
}.  
Figure 4 compares the 6.2 $\mu$m and 11.3 $\mu$m PAH
luminosities with the observed infrared luminosities.

The observed L$_{\rm 6.2PAH}$/L$_{\rm IR}$ ratios are (0.3--2.4) $\times$
10$^{-3}$ (Table 3), or 9--71 \% of 3.4 $\times$ 10$^{-3}$ for 
modestly obscured starburst galaxies.  
In the majority of the observed ULIRGs, the ratios are $<$1.7 $\times$
10$^{-3}$, or $<$50 \% of 3.4 $\times$ 10$^{-3}$ (see also Figure 4). 
Taken at face value, the modestly obscured starbursts detected in these
ULIRGs account for 9--71 \% (mostly $<$50 \%) of their infrared
luminosities.
The same argument can be applied to the observed 
L$_{\rm 11.3PAH}$/L$_{\rm IR}$ ratios for the ULIRGs. 
The L$_{\rm 11.3PAH}$/L$_{\rm IR}$ ratios are 
(0.3--1.8) $\times$ 10$^{-3}$ (Table 3), or 21--100 \% of 
1.4 $\times$ 10$^{-3}$ for modestly obscured starburst
galaxies.
On the basis of the L$_{\rm 11.3PAH}$/L$_{\rm IR}$ ratios, 
there are some ULIRGs whose infrared luminosities can be explained 
by the detected modestly obscured (A$_{\rm V}$ $<$ 20 mag) starbursts.
However, for ULIRGs with $<$0.7 $\times$ 10$^{-3}$ ($<$50 \% of 1.4
$\times$ 10$^{-3}$), additional energy sources would be required.
The 11.3 $\mu$m PAH emission tends to provide higher contributions 
from modestly obscured starbursts to the infrared luminosities of
ULIRGs, than the 6.2 $\mu$m PAH emission (Figure 4), possibly because
the 11.3 $\mu$m PAH emission originates in larger-sized PAHs which may
be less susceptible to destruction by the energetic radiation from AGNs
and intense starbursts in ULIRGs \citep{smi07}.
Luminosities that are not accounted for by the detected
modestly obscured starbursts might originate from 
(1) highly obscured (A$_{\rm V}$ $>>$ 20 mag) starbursts, in which 
the PAH emission flux is substantially attenuated by dust  
extinction, and/or (2) buried AGNs that produce strong infrared
radiation, but virtually no PAH emission.

We note that emission from {\it extreme} starbursts, which consist of
HII-regions only, without PDRs and molecular gas, can be PAH-free, 
because PAH molecules can be destroyed inside the HII-regions 
themselves \citep{sel81}. 
If starbursts are exceptionally more centrally concentrated 
than the surrounding molecular gas and dust (Figure 1e of Imanishi 
et al. 2007), such HII-region-only starbusts could happen. 
However, in the case of ULIRG's cores, such {\it extreme} starbursts 
require extremely high emission surface brightnesses, and so are 
unlikely \citep{ima07a}, if not completely ruled out.  
We thus use the term ``buried AGN signatures'', rather than 
``buried AGN evidence'' throughout this manuscript.

\subsection{ULIRGs that could contain luminous buried AGNs}

\subsubsection{Low equivalent width PAH emission}

Highly-obscured normal starbursts and buried AGNs are basically
distinguishable, based on the equivalent width of the PAH emission.
A PAH equivalent width (EW$_{\rm PAH}$) must always be large
in a normal starburst (with PDRs) with a mixed dust/source geometry,
regardless of the amount of dust extinction, because both the PAH and
nearby continuum emission are similarly flux-attenuated. 
Thus, a small EW$_{\rm PAH}$ value suggests a contribution from
a PAH-free continuum-emitting energy source, namely an AGN
\citep{ima07a,ima09}. 

Following \citet{ima07a} and \citet{ima09}, we classify ULIRGs with
EW$_{\rm 6.2PAH}$  $<$ 180 nm and EW$_{\rm 11.3PAH}$ $<$ 200 nm as
sources displaying clear signatures of luminous AGNs, because a
considerable contribution from a PAH-free continuum would be required
for these sources.
Figure 5 is a plot of the distribution of EW$_{\rm 6.2PAH}$ and 
EW$_{\rm 11.3PAH}$ values.
Table 6 (columns 2--3) records the detection or non-detection of buried
AGN signatures in terms of the PAH equivalent width threshold. 
The EW$_{\rm 6.2PAH}$ method provides a much larger buried AGN fraction
(7/17; 41\%) than the EW$_{\rm 11.3PAH}$ method (2/17; 12\%), as seen in
\citet{ima07a} and \citet{ima09}.  
An explanation for this is that the 11.3 $\mu$m PAH emission
feature is inside the strong 9.7 $\mu$m silicate dust absorption
feature. Thus, the buried AGN continuum emission at $\lambda_{\rm rest}$
$\sim$ 11.3 $\mu$m is severely attenuated, and has little effect on the
equivalent widths of the 11.3 $\mu$m PAH emission from less obscured
starburst regions outside the AGNs \citep{ima07a,ima09}.

\subsubsection{Optical depths of dust absorption features}

Based on the low EW$_{\rm PAH}$ method, a buried AGN with very weak
starbursts is easily detectable.
If strong starburst activity is present, AGN detection becomes difficult,
but a {\it weakly obscured} AGN is still detectable, because dilution of
the PAH emission by the AGN's PAH-free continuum can be significant. 
However, a {\it deeply buried} AGN with strong starbursts is very
difficult to detect. Even if the intrinsic luminosity of a buried AGN
is large, the AGN flux will be more highly attenuated by dust extinction
than the surrounding starburst emission, keeping the observed 
EW$_{\rm PAH}$ value quite large.

To determine the presence of a deeply buried AGN with
strong starbursts, we use the optical depths of the silicate dust
absorption features. Specifically, these values can be used to
distinguish whether the energy sources are spatially well mixed with
dust (a normal starburst),  
or are more centrally concentrated than the dust (a buried AGN) ($\S$1).
\citet{ima07a} obtained a maximum value of $\tau_{9.7}'$ $=$ 1.7 for
a normal starburst with a mixed dust/source geometry.
Considering the possible uncertainties of the $\tau_{9.7}'$
estimate, coming from continuum determination ambiguities and
statistical errors of individual data points, we classify ULIRGs with
$\tau_{9.7}'$ 
$\geq$ 2 as potential locations of luminous, but deeply buried AGNs with
centrally concentrated energy source geometries.
Five ULIRGs have $\tau_{9.7}'$ $\geq$ 2 (Tables 4 and 6).

As noted by \citet{ima09}, although this large $\tau_{9.7}$ method is
sensitive to deeply buried AGNs, it is obviously insensitive to 
weakly obscured AGNs, which are better probed with the low EW$_{\rm PAH}$
method. Thus, the low EW$_{\rm PAH}$ and large $\tau_{9.7}$ methods
are complementary to each other. 
At the same time, a normal starburst nucleus with a mixed dust/source
geometry and a large amount of {\it foreground screen dust in an edge-on
host galaxy} (Figure 1d of Imanishi et al. 2007), and 
an exceptionally centrally-concentrated starburst (Figure 1e of Imanishi
et al. 2007), are capable of producing large $\tau_{9.7}'$ values. 
However, it is unlikely that the majority of the ULIRGs with $\tau_{9.7}'$ 
$\geq$ 2 fit these non-AGN cases \citep{ima07a,ima09}.

\subsubsection{Strong dust temperature gradients}

As explained by \citet{ima07a}, a buried AGN with a centrally concentrated
energy source geometry should have a strong dust temperature gradient,
because the inner dust (close to the central energy source) has a higher
temperature than the outer dust, whereas a normal starburst nucleus with
a mixed dust/source geometry does not exhibit this behavior. 
The presence of a strong dust temperature gradient could be detected by
comparing the optical depths of the 9.7 $\mu$m and 18 $\mu$m silicate dust
absorption features, as long as the contaminations from weakly obscured
starbursts to observed infrared fluxes are not severe. 
$\tau_{18}'$/$\tau_{9.7}'$ $<$ 0.3 could be taken as the signature of a
strong dust temperature gradient \citep{ima07a,ima09}.  
This could provide an additional signature for a buried AGN.
Although this was the case for 12 sources studied by \citet{ima07a} and 
\citet{ima09}, no such case was found among the 17 newly observed
ULIRGs (Table 4, column 4).  
Because of the contamination from weakly obscured starburst activity, as
well as other possible ambiguities \citep{ima07a}, not all buried AGNs
actually satisfy $\tau_{18}'$/$\tau_{9.7}'$ $<$ 0.3.  
The criterion $\tau_{18}'$/$\tau_{9.7}'$ $<$ 0.3 is a
sufficient condition for an additional signature of a buried AGN,
but it is not a necessary condition.

\subsubsection{Combination of energy diagnostic methods}

Table 6 (column 5) summarizes the strengths of the detected buried AGN
signatures in the {\it Spitzer} IRS 5--35 $\mu$m spectra, based on the
(1) low EW$_{\rm PAH}$ and (2) large $\tau_{9.7}'$ methods.
If buried AGN signatures are found by both methods, or by the first method,
the ULIRGs are classified as {\it strong} buried AGN candidates, marked
with open circles. 
Eight ULIRGs show strong signatures of luminous buried AGNs (Table 6).

The detection rate of luminous buried AGNs in the newly observed
optically non-Seyfert ULIRGs (8/17; 47\%) is significantly smaller than
the rate obtained by Imanishi (2009) (14/20; 70\%). 
This may be partly due to the smaller fraction of ULIRGs with 
L$_{\rm IR}$ $\geq$ 10$^{12.3}$L$_{\odot}$ (6/17; 35\%) in this sample,
compared with \citet{ima09} (17/20; 85\%).

\subsubsection{Extinction-corrected intrinsic buried AGN luminosities}

For ULIRGs with very low PAH equivalent widths, the observed fluxes are
taken to be mostly the result of AGN-heated, PAH-free dust continuum
emission.   
We can estimate the extinction-corrected intrinsic dust emission luminosity 
at $\sim$10 $\mu$m ($\nu$F$_\nu$), heated by the AGN, on the
basis of the observed fluxes at $\lambda_{\rm rest}$ $\sim$ 10 $\mu$m and 
dust extinction toward the 10 $\mu$m continuum emitting regions inferred
from $\tau_{9.7}'$ \citep{ima07a,ima09}.  
As argued by \citet{ima07a} and \citet{ima09}, in a pure AGN 
with a simple spherical dust distribution, dust emission luminosity is
conserved at each temperature from hot inside regions to cool outside
regions (Figure 2 of Imanishi et al. 2007).  
The extinction-corrected 10 $\mu$m luminosity ($\nu$F$_\nu$) should be
comparable to the intrinsic luminosity of the AGN's primary radiation.

To obtain the estimate, we followed \citet{ima07a} and \citet{ima09}.
The flux attenuation of the 8 or 13 $\mu$m continuum outside 
the 9.7 $\mu$m silicate feature is 10$^{\tau_{9.7}'/2.3/2.5}$
\citep{rie85}, and ranges from a factor of 
1.3 (IRAS 14202+2615; $\tau_{9.7}'$ $\sim$ 0.7) to 3.3 (IRAS 04074$-$2801;  
$\tau_{9.7}'$ $\sim$ 3.0). We found that the extinction-corrected
intrinsic buried AGN luminosities for selected ULIRGs with very low
EW$_{\rm PAH}$ values are a few $\times$ 10$^{45}$ ergs s$^{-1}$, or
$\sim$10$^{12}$L$_{\odot}$ in the maximum case (Table 7).
Figure 6 compares the intrinsic buried AGN luminosity and the infrared
luminosity.
The detected buried AGNs could explain a significant, but not dominant
fraction (6--33\%), of the luminosities of these ULIRGs.  
In Table 7, the luminosities of the detected modestly obscured starbursts,
estimated from the L$_{\rm 6.2PAH}$ and L$_{\rm 11.3PAH}$ values, 
are also listed for the sake of comparison.

Our method to estimate the intrinsic buried AGN luminosities is 
very simple and straightforward, with a small amount of free parameters. 
The largest possible uncertainty is dust extinction curve in ULIRGs,
which might be different from that established in the Galactic
inter-stellar medium which we adopted.
Different dust grain size distribution can produce different 
dust extinction ratio between widely separate wavelengths, such as 
optical and 10 $\mu$m. 
However, as mentioned by Imanishi et al. (2007; their Appendix), 
our estimate is dependent only on the dust extinction ratio between 
8 or 13 $\mu$m continuum and $\tau_{9.7}'$, which should be insensitive 
to dust size distribution, because of the proximity of the wavelengths.
Hence, our method is robust to this possible uncertainty.

\subsection{Buried AGN fraction as a function of galaxy infrared
luminosity} 

\citet{ima07a} investigated the presence of luminous buried AGNs 
in 48 ULIRGs at $z <$ 0.15, classified optically as non-Seyferts (28
LINER and 20 HII-region ULIRGs).
\citet{ima09} did the same in 20 ULIRGs at $z >$ 0.15, classified
optically as non-Seyferts (10 LINER, 6 HII-region, 4 unclassified
ULIRGs).
In the present research, we have continued this work in 13 optically
non-Seyfert ULIRGs at $z >$ 0.15  (5 LINER, 6 HII-region, 2 unclassified
ULIRGs) and four optically unclassified ULIRGs at $z <$ 0.15. 
Taken together, these 85 observed sources comprise a 
{\it complete} sample of optically non-Seyfert ULIRGs in the {\it IRAS}
1 Jy sample. 

Following \citet{ima09}, we divide the observed ULIRGs into those with
10$^{12}$L$_{\odot}$ $\leq$ L$_{\rm IR}$ $<$ 10$^{12.3}$L$_{\odot}$ 
and those with L$_{\rm IR}$ $\geq$ 10$^{12.3}$L$_{\odot}$.  
After combining the results of \citet{ima07a}, \citet{ima09} and this paper, 
we summarize the detectable buried AGN fraction in Table 8.
Using this complete sample, we confirm the previously discovered trend
\citep{ima08,ima09,vei09,nar09} that the detectable buried AGN
fraction is significantly higher in ULIRGs with    
L$_{\rm IR}$ $\geq$ 10$^{12.3}$L$_{\odot}$ (22/31; 71\%) than in those with
10$^{12}$L$_{\odot}$ $\leq$ L$_{\rm IR}$ $<$ 10$^{12.3}$L$_{\odot}$
(15/54; 28\%).   
Figure 7 illustrates this trend by including galaxies with 
L$_{\rm IR}$ $<$ 10$^{12}$L$_{\odot}$.
Figure 5 displays the distributions of EW$_{\rm 6.2PAH}$, EW$_{\rm 11.3PAH}$ 
and $\tau_{9.7}'$ as functions of galaxy infrared luminosity. 
It is clear that the fraction of galaxies that meet the
requirements for a buried AGN increases with increasing galaxy infrared
luminosity in all plots.
 
In Table 8, we also find that buried AGNs are more important 
in LINER ULIRGs (21/43; 49\%) than in HII-region ULIRGs (12/32; 38\%), 
as has been previously suggested \citep{idm06,ima07a,ima08,ima09}.
However, the difference is now small in this complete sample, and is
possibly caused by a higher fraction of ULIRGs with L$_{\rm IR}$ $\geq$
10$^{12.3}$L$_{\odot}$ in the LINER sample (18/43; 42\%) than in the 
HII-region sample (10/32; 31\%) (Table 8).
The buried AGN fraction is also higher in ULIRGs at $z >$ 0.15 than in
those at $z \leq$ 0.15, but this may be due to the presence of a 
larger fraction of more luminous ULIRGs in a more distant sample.

For an AGN surrounded by dust with a torus-shaped distribution, the
visibility of  
the central AGN is expected to increase with increasing AGN luminosity,
because the innermost dust sublimation radius of the torus is
proportional to the square root of the central AGN luminosity, and thus 
(assuming a constant torus scale height) the opening angle of the AGN is
greater in luminous AGNs (the so-called receding torus model:
Lawrence 1991; Simpson 2005; Arshakian 2005). 
In this scenario, the fraction of optically elusive buried AGNs 
is not large in luminous AGNs. 
However, this model is applicable {\it only if} the total amount of dust
surrounding the central AGN does not vary significantly, depending on 
the galaxy and AGN luminosity. 
ULIRGs are driven by mergers of gas-rich galaxies, and gas/dust is
quickly transported to the nuclear regions by the gravitational
torques, producing a highly-concentrated nuclear gas/dust distribution 
\citep{hop06}.
AGNs can be easily buried by the increasing amount
of nuclear gas/dust surrounding the central AGNs, and thus we believe
that a high buried AGN fraction is a natural state of affairs in ULIRGs.

The higher fractions of buried AGNs (found in this work) and optical 
Seyferts \citep{vei99,got05} indicate that AGNs become intrinsically more 
important with increasing galaxy infrared luminosity.  
Namely, {\it the AGN-starburst connections depend on galaxy luminosity}. 
This is the primary consequence of our results. 

Figure 6 suggests that AGN's energetic contributions are generally
10--50\%, or $\sim$30\% in median value, in ULIRGs with buried AGN
signatures, which roughly agrees with other independent estimates 
\citep{vei09,nar09} in overall ULIRG sample.  
If 30\% of the infrared luminosity comes from a buried AGN in a ULIRG 
with L$_{\rm IR}$ = 10$^{12.3}$L$_{\odot}$ (= 2 $\times$
10$^{12}$L$_{\odot}$), then the remaining infrared luminosity,
originating in starbursts (including highly obscured A$_{\rm V}$ $>>$ 20
mag ones), is 1.4 $\times$ 10$^{12}$L$_{\odot}$. 
In the meanwhile, buried AGN contribution is generally insignificant in
galaxies with L$_{\rm IR}$ = 10$^{11.3}$L$_{\odot}$ (= 2 $\times$ 
10$^{11}$L$_{\odot}$) (Figure 7), and in this case, the
starburst-originating luminosity is $\sim$2 $\times$
10$^{11}$L$_{\odot}$. 
Namely, in ULIRGs, the {\it relative} energetic importance of buried AGN
is higher, but the {\it absolute} luminosities of starbursts, and
thereby the star-formation rates, are also higher than galaxies with
lower infrared luminosities.  
Unless the starburst duration time is drastically different, 
ULIRGs will produce more stars in the future, and evolve into 
more massive galaxies with larger stellar masses, than less 
infrared luminous galaxies. 

Given the mass correlation between SMBH and spheroidal stellar component
\citep{mag98,fer00}, galaxies with lower infrared luminosities should
also contain SMBHs whose masses are proportional to those of
spheroidal stars.  
The higher AGN contributions in ULIRGs suggest that SMBHs in ULIRGs 
(progenitors of massive galaxies) are actively
mass-accreting, and consequently, can have stronger radiation feedbacks 
to the surrounding gas and dust than mildly mass accreting SMBHs in less
infrared luminous galaxies.
A related phenomenon may be galaxy downsizing, where the more massive
galaxies (with currently larger stellar masses) completed 
their major star formation during an earlier cosmic age \citep{cow96,bun05}.   
It has been suggested that in these more massive galaxies, AGN
feedback was stronger in the past, heating or expelling gas 
and suppressing further star formation over a shorter time scale 
\citep{gra04,bow06,sij07}.
Buried AGNs surrounded by a large amount of gas and dust may have
particularly strong feedback, compared to already visible AGNs
with a relatively thin dust covering, and thus are the important population 
for determining the interplay between AGNs and host galaxies.

Figure 8 illustrates our findings that {\it buried AGNs are relatively 
more important energetically in galaxies that are currently more infrared
luminous, and in the future, these galaxies will evolve into more
massive galaxies with larger stellar masses.} 
The energetic contributions from buried AGNs become discernible in ULIRGs,
which are thought to evolve into galaxies with stellar masses of several
$\times$ 10$^{10}$M$_{\odot}$, based on infrared velocity dispersion
measurements of ULIRG's host galaxies \citep{das06}.  
A similar eventual spheroidal stellar mass is derived for ULIRGs with
detectable buried AGN signatures, from the intrinsic buried AGN luminosities
\citep{ima09}, if we assume that AGN luminosities are Eddington limits
and if the widely accepted mass correlation between SMBHs and spheroidal
steller components holds \citep{mag98,fer00}. 
This mass with several $\times$ 10$^{10}$M$_{\odot}$ roughly
corresponds to the stellar mass that separates red, massive galaxies
with low current star formation rates (major star formation has already
been completed) and blue, less massive galaxies with ongoing active star
formation in the local universe \citep{kau03}. 
Thus, our results may offer support to the AGN feedback
scenario as the origin of the galaxy downsizing phenomenon.

\section{Summary}

We have presented the results of {\it Spitzer} IRS infrared 5--35 $\mu$m
low-resolution (R $\sim$ 100) spectroscopy of 17 nearby ULIRGs at $z <$
0.2, optically classified as non-Seyferts (LINERs, HII-regions, and 
unclassified, i.e., no optical AGN signatures). 
Optically elusive, but intrinsically luminous buried AGNs were 
searched for in these optically non-Seyfert ULIRGs, on the basis of
the strengths of PAH emission and silicate dust absorption features.  
We then combined these results with those of our previous studies of
nearby ULIRGs, 
using {\it Spitzer} IRS, to investigate the energetic importance of
buried AGNs in a {\it complete sample} of optically non-Seyfert ULIRGs
in the local universe at $z <$ 0.3 (85 sources altogether). 
We arrived at the following primary conclusions. 

\begin{enumerate}

\item Among the 17 newly observed optically non-Seyfert ULIRGs, 
the signatures of important energy contributions from buried AGNs 
were found in eight sources. 
In these sources, the extinction-corrected intrinsic buried AGN
luminosities were estimated at up to 
$\sim$10$^{12}$L$_{\odot}$, accounting for a
significant fraction (6--33\%) of the observed infrared luminosities of
these ULIRGs. 

\item By combining our new results with those of our previous studies
\citep{ima07a,ima09}, we found that buried AGNs are energetically
important in 37 sources of the complete ULIRG sample of 85 
(37/85 = 44\%), confirming previous suggestion that optically elusive,
luminous buried AGNs are common in the ULIRGs of the local universe. 

\item We investigated the fraction of detectable luminous buried AGNs
by separating ULIRGs with 10$^{12}$L$_{\odot}$ $\leq$ L$_{\rm IR}$ $<$ 
10$^{12.3}$L$_{\odot}$ (54 sources) and L$_{\rm IR}$ $\geq$
10$^{12.3}$L$_{\odot}$ (31 sources).
We found that luminous buried AGNs were much more common in the latter
ULIRGs (22/31 = 71\%) than in the former ULIRGs (15/54 = 28\%), confirming
the previous arguments that buried AGNs become more
energetically important with increasing galaxy infrared luminosity. 

\item Given the higher fraction of optical Seyferts 
(optically identified AGNs) in ULIRGs with higher infrared
luminosities, luminous AGNs are more common 
in ULIRGs with L$_{\rm IR}$ $\geq$ 10$^{12.3}$L$_{\odot}$ than 
in ULIRGs with 10$^{12}$L$_{\odot}$ $\leq$ L$_{\rm IR}$ $<$ 
10$^{12.3}$L$_{\odot}$.
Because the detection rate of both optically identified Seyfert AGNs and
optically elusive buried AGNs is substantially lower in galaxies 
with lower infrared luminosities (L$_{\rm IR}$ $<$
10$^{12}$L$_{\odot}$), we can conclude that the energetic importance of
AGNs increases with increasing galaxy infrared luminosity, suggesting
that {\it the AGN-starburst connections are luminosity dependent}. 
This may be related to the widely-proposed AGN feedback scenario for 
the galaxy downsizing phenomenon.

\end{enumerate}

\acknowledgments

This work is based on observations made with the 
Spitzer Space Telescope, operated by the Jet Propulsion
Laboratory at California Institute of Technology, under a contract with
NASA. Support for this work was provided by NASA, and also by an award
issued by JPL/Caltech.  
We thank the anonymous referee for his/her valuable comments which 
help significantly improve the clarity of the arguments in this 
manuscript. 
M.I. is supported by Grants-in-Aid for Scientific Research
(19740109). 
R.M. acknowledges partial support from INAF and ASI, through
contract ASI-INAF I/016/07/0.
This research made use of the SIMBAD database, operated at
CDS, Strasbourg, France, and the NASA/IPAC Extragalactic Database
(NED), which is operated by the Jet Propulsion Laboratory at California
Institute of Technology, under a contract with NASA.

\clearpage

\begin{deluxetable}{lcrrrrcrc}
\tabletypesize{\scriptsize}
\tablecaption{Observed ULIRGs and their {\it IRAS}-based infrared
emission properties 
\label{tbl-1}}
\tablewidth{0pt}
\tablehead{
\colhead{Object} & \colhead{Redshift}   & 
\colhead{f$_{\rm 12}$}   & 
\colhead{f$_{\rm 25}$}   & 
\colhead{f$_{\rm 60}$}   & 
\colhead{f$_{\rm 100}$}  & 
\colhead{log L$_{\rm IR}$} & 
\colhead{f$_{25}$/f$_{60}$} & 
\colhead{Optical}   \\
\colhead{} & \colhead{}   & \colhead{(Jy)} & \colhead{(Jy)} 
& \colhead{(Jy)} & \colhead{(Jy)}  & \colhead{L$_{\odot}$} & \colhead{}
& \colhead{Class}   \\
\colhead{(1)} & \colhead{(2)} & \colhead{(3)} & \colhead{(4)} & 
\colhead{(5)} & \colhead{(6)} & \colhead{(7)} & \colhead{(8)} & 
\colhead{(9)}
}
\startdata
IRAS 04074$-$2801 & 0.153 & $<$0.07 & 0.07 & 1.33 & 1.72 & 12.2 & 0.05 (C) & LINER \\  
IRAS 05020$-$2941 & 0.154 & $<$0.06 & 0.10 & 1.93 & 2.06 & 12.3 & 0.05 (C) & LINER \\  
IRAS 13106$-$0922 & 0.174 & $<$0.12 & $<$0.06 & 1.24 & 1.89 & 12.3 & $<$0.05 (C) & LINER \\  
IRAS 14121$-$0126 & 0.151 & 0.06    & 0.11 & 1.39 & 2.07 & 12.3 & 0.08 (C) & LINER \\  
IRAS 21477+0502   & 0.171 & $<$0.09 & 0.16 & 1.14 & 1.46 & 12.3 & 0.14 (C) & LINER \\  
IRAS 03209$-$0806 & 0.166 & $<$0.10 & $<$0.13 & 1.00 & 1.69 & 12.2 & $<$0.13 (C) & HII-region \\  
IRAS 10594+3818   & 0.158 & $<$0.09 & $<$0.15 & 1.29 & 1.89 & 12.2 & $<$0.12 (C) & HII-region \\  
IRAS 12447+3721   & 0.158 & $<$0.12 & 0.10 & 1.04 & 0.84 & 12.1 & 0.10 (C) & HII-region \\  
IRAS 14202+2615   & 0.159 & 0.18    & 0.15 & 1.49 & 1.99 & 12.4 & 0.10 (C) & HII-region \\  
IRAS 15043+5754   & 0.151 & $<$0.12 & 0.07 & 1.02 & 1.50 & 12.1 & 0.07 (C) & HII-region \\  
IRAS 22088$-$1831 & 0.170 & $<$0.09 & 0.07 & 1.73 & 1.73 & 12.4 & 0.04 (C) & HII-region \\  
IRAS 02480$-$3745 & 0.165 & $<$0.05 & $<$0.11 & 1.25 & 1.49 & 12.2 & $<$0.09 (C) & unclassified \\  
IRAS 08591+5248   & 0.158 & $<$0.10 & $<$0.16 & 1.01 & 1.53 & 12.2 & $<$0.16 (C) & unclassified \\  
IRAS 02021$-$2103 & 0.116 & $<$0.07 & 0.30 & 1.45 & 1.72 & 12.0 & 0.21 (W) & unclassified \\  
IRAS 08474+1813   & 0.145 & $<$0.10 & $<$0.19 & 1.28 & 1.54 & 12.1 & $<$0.15 (C) & unclassified \\
IRAS 14197+0813   & 0.131 & $<$0.17 & $<$0.19 & 1.10 & 1.66 & 12.0 & $<$0.18 (C) & unclassified \\  
IRAS 14485$-$2434 & 0.148 & $<$0.11 & $<$0.15 & 1.02 & 1.05 & 12.1 & $<$0.15 (C) & unclassified \\ \hline
\enddata

\tablecomments{
Col.(1): Object name.  
Col.(2): Redshift.
Col.(3)--(6): f$_{12}$, f$_{25}$, f$_{60}$, and f$_{100}$ are 
{\it IRAS} fluxes at 12 $\mu$m, 25 $\mu$m, 60 $\mu$m, and 100 $\mu$m in
[Jy], respectively, taken from \citet{kim98}.
Col.(7): Decimal logarithm of infrared (8$-$1000 $\mu$m) luminosity
in units of solar luminosity (L$_{\odot}$), calculated with
$L_{\rm IR} = 2.1 \times 10^{39} \times$ D(Mpc)$^{2}$
$\times$ (13.48 $\times$ $f_{12}$ + 5.16 $\times$ $f_{25}$ +
$2.58 \times f_{60} + f_{100}$) ergs s$^{-1}$ \citep{sam96}.
Because the calculation is based on our adopted cosmology, the infrared
luminosities differ slightly ($<$10\%) from the values shown in Kim \&
Sanders (1998, Table 1, column 15).  
For sources with upper limits in some {\it IRAS} band, 
we can derive the respective upper and lower limits for infrared luminosity,
assuming that the actual flux is between the {\it IRAS}-upper limit and zero.
The difference in the upper and lower values is usually very small, less
than 0.25 dex.
We assume that the infrared luminosity is the average of these values. 
Col.(8): {\it IRAS} 25 $\mu$m to 60 $\mu$m flux ratio.
ULIRGs with f$_{25}$/f$_{60}$ $<$ 0.2 and $>$ 0.2 are
classified as cool and warm sources (denoted as ``C'' and ``W''),
respectively \citep{san88b}.
Col.(9): Optical spectral classification by \citet{vei99}.
}

\end{deluxetable}

\begin{deluxetable}{lclcccc}
\tabletypesize{\small}
\tablecaption{{\it Spitzer} IRS observing log
\label{tbl-2}}
\tablewidth{0pt}
\tablehead{
\colhead{Object} & \colhead{PID} &\colhead{Date} & \multicolumn{4}{c}
{Integration time [sec]} \\ 
\colhead{} & \colhead{} & \colhead{[UT]} & \colhead{SL2} & \colhead{SL1} &
\colhead{LL2} & \colhead{LL1} \\ 
\colhead{(1)} & \colhead{(2)} & \colhead{(3)} & \colhead{(4)} & 
\colhead{(5)} & \colhead{(6)} & \colhead{(7)} 
}
\startdata 
IRAS 04074$-$2801 & 50008 & 2009 Mar 8 & 480 & 480 & 240 & 240 \\ 
IRAS 05020$-$2941 & 50008 & 2008 Dec 5 & 240 & 240 & 240 & 240 \\
IRAS 13106$-$0922 & 50008 & 2009 Mar 6 & 480 & 480 & 240 & 240 \\ 
IRAS 14121$-$0126 & 50008 & 2009 Mar 5 & 240 & 240 & 240 & 240 \\
IRAS 21477+0502 \tablenotemark{a} & 50008 & 2008 Dec 5 & 960 & 960 & 480 & 480 \\ 
IRAS 03209$-$0806 & 50008 & 2009 Mar 8 & 480 & 480 & 240 & 240 \\
IRAS 10594+3818 \tablenotemark{b} & 50008 & 2009 Jan 11, 15 & 960 & 960 & 480 & 480 \\
IRAS 12447+3721   & 50008 & 2009 Jan 24 & 240 & 240 & 240 & 240 \\
IRAS 14202+2615 \tablenotemark{c}  & 50008 & 2009 Mar 3, 2009 Apr 2 & 480 & 480 & 480 & 480 \\
IRAS 15043+5754  \tablenotemark{d} & 50008 & 2009 Feb 26 & 960 & 960 & 480 & 480 \\
IRAS 22088$-$1831 \tablenotemark{e} & 50008 & 2008 Dec 5 & 960 & 960 & 480 & 480 \\
IRAS 02480$-$3745 & 50008 & 2009 Jan 25 & 240 & 240 & 240 & 240 \\
IRAS 08591+5248   & 50008 & 2008 Dec 9 & 240 & 240 & 240 & 240 \\
IRAS 02021$-$2103 & 3187 + 50008 & 2005 Jan 15, 2009 Jan 15 & 240 & 240 & 240 & 240 \\
IRAS 08474+1813   & 30407 & 2007 Dec 5 & 168 & 168 & 180 & 180 \\
IRAS 14197+0813   & 3187 + 50008 & 2005 Feb 13, 2009 Mar 3 & 240 & 240 & 240 & 240 \\
IRAS 14485$-$2434 & 50008 & 2009 Apr 6  & 240 & 240 & 240 & 240 \\ \hline
\enddata

\tablenotetext{a}{Although separate spectroscopy of E and W nuclei
with a separation of $\sim$7.5 arcsec \citep{kim02} had been proposed,
the E nucleus was observed.}  

\tablenotetext{b}{Although separate spectroscopy of SW and NE nuclei
with a separation of $\sim$2 arcsec \citep{kim02} had been proposed, the
SW nucleus was observed.}  

\tablenotetext{c}{Although separate spectroscopy of SE and NW nuclei
with a separation of $\sim$6 arcsec \citep{kim02} had been proposed, the
SE nucleus was observed.}  

\tablenotetext{d}{Although separate spectroscopy of S and N nuclei
with a separation of $\sim$2.5 arcsec \citep{kim02} had been proposed,
the S nucleus was observed.}  

\tablenotetext{e}{Although separate spectroscopy of E and W nuclei
with a separation of $\sim$2 arcsec \citep{kim02} had been proposed, the
E-nucleus was observed.}  

\tablecomments{
Col.(1): Object name.
Col.(2): PID number: 50008 (PI: Imanishi), 3187 (PI: Veilleux), and
30407 (PI: Darling). 
Col.(3): Observing date in UT. 
Col.(4): Net on-source integration time for SL2 spectroscopy in [sec].
Col.(5): Net on-source integration time for SL1 spectroscopy in [sec].
Col.(6): Net on-source integration time for LL2 spectroscopy in [sec]
Col.(7): Net on-source integration time for LL1 spectroscopy in [sec].
}

\end{deluxetable}

\begin{deluxetable}{lccccccccc}
\tabletypesize{\scriptsize}
\tablecaption{Observed properties of PAH emission features 
\label{tbl-3}}
\tablewidth{0pt}
\tablehead{
\colhead{Object} & \colhead{EW$_{\rm 6.2PAH}$} & 
\colhead{EW$_{\rm 7.7PAH}$ \tablenotemark{a}} & 
\colhead{EW$_{\rm 11.3PAH}$}  & \colhead{L$_{\rm 6.2PAH}$} & 
\colhead{L$_{\rm 7.7PAH}$ \tablenotemark{a}} &
\colhead{L$_{\rm 11.3PAH}$} & \colhead{L$_{\rm 6.2PAH}$/L$_{\rm IR}$} & 
\colhead{L$_{\rm 11.3PAH}$/L$_{\rm IR}$} \\
\colhead{} & \colhead{[nm]} & \colhead{[nm]} & \colhead{[nm]} &
\colhead{10$^{42}$ [ergs s$^{-1}$]} & \colhead{10$^{42}$ [ergs s$^{-1}$]} & 
\colhead{10$^{42}$ [ergs s$^{-1}$]} & \colhead{[$\times$ 10$^{-3}$]} & 
\colhead{[$\times$ 10$^{-3}$]} \\  
\colhead{(1)} & \colhead{(2)} & \colhead{(3)} & \colhead{(4)} &
\colhead{(5)} & \colhead{(6)} & \colhead{(7)} & \colhead{(8)} & 
\colhead{(9)} 
}
\startdata 
IRAS 04074$-$2801 &  60 & 420 & 235 & 3.3 & 32.8 & 2.5 & 0.6 & 0.4 \\ 
IRAS 05020$-$2941 & 130 & 605 & 330 & 7.0 & 47.1 & 4.2 & 0.9 & 0.5 \\
IRAS 13106$-$0922 & 115 & 425 & 565 & 3.3 & 27.6 & 3.5 & 0.4 & 0.4 \\ 
IRAS 14121$-$0126 & 270 & 635 & 345 & 11.9 & 38.2 & 7.8 & 1.7 & 1.2 \\
IRAS 21477+0502   & 235 & 575 & 185 & 7.4 & 24.2 & 5.4 & 1.0 & 0.7 \\ 
IRAS 03209$-$0806 & 285 & 555 & 410 & 9.9 & 26.4 & 9.2 & 1.6 & 1.5 \\
IRAS 10594+3818   & 350 & 780 & 490 & 16.0 & 46.0 & 11.1 & 2.4 & 1.7 \\
IRAS 12447+3721   & 210 & 625 & 260 & 7.0 & 24.0 & 4.1 & 1.4 & 0.8 \\
IRAS 14202+2615   & 160 & 455 & 265 & 18.7 & 52.9 & 13.6 & 1.9 & 1.4 \\
IRAS 15043+5754   & 285 & 770 & 555 & 8.0 & 29.3 & 6.7 & 1.6 & 1.3 \\
IRAS 22088$-$1831 &  90 & 455 & 195 & 3.0 & 23.6 & 2.3 & 0.3 & 0.3 \\
IRAS 02480$-$3745 & 325 & 915 & 535 & 7.3 & 25.6 & 5.0 & 1.2 & 0.8 \\
IRAS 08591+5248   & 310 & 695 & 535 & 9.7 & 30.7 & 9.7 & 1.8 & 1.8 \\
IRAS 02021$-$2103 & 285 & 490 & 275 & 5.5 & 13.0 & 4.2 & 1.4 & 1.0 \\
IRAS 08474+1813   & 170 & 985 & 285 & 2.2 & 13.1 & 1.4 & 0.4 & 0.3 \\
IRAS 14197+0813   & 305 & 565 & 325 & 3.8 & 11.4 & 3.7 & 0.9 & 0.9 \\
IRAS 14485$-$2434 & 150 & 495 & 295 & 6.8 & 26.2 & 6.8 & 1.6 & 1.6 \\ \hline
\enddata

\tablenotetext{a}{We consider the flux excess at $\lambda_{\rm rest}$ =
7.3--8.1 $\mu$m above an adopted continuum level to be 7.7 $\mu$m PAH 
emission, to reduce the effects of the strong 9.7 $\mu$m silicate dust
absorption feature. 
The 7.7 $\mu$m PAH emission strengths may be significantly different
from those estimated by other authors.}

\tablecomments{
Col.(1): Object name.  
Col.(2): Rest-frame equivalent width of the 6.2 $\mu$m PAH emission in
[nm]. 
Col.(3): Rest-frame equivalent width of the 7.7 $\mu$m PAH emission in
[nm]. 
Col.(4): Rest-frame equivalent width of the 11.3 $\mu$m PAH emission in
[nm]. 
Col.(5): Luminosity of the 6.2 $\mu$m PAH emission in units 
of 10$^{42}$ [ergs s$^{-1}$].
Col.(6): Luminosity of the 7.7 $\mu$m PAH emission in units 
of 10$^{42}$ [ergs s$^{-1}$].
Col.(7): Luminosity of the 11.3 $\mu$m PAH emission in units 
of 10$^{42}$ [ergs s$^{-1}$].
Col.(8): The 6.2 $\mu$m PAH to infrared luminosity ratio in units of 
10$^{-3}$. 
The ratio for normal starbursts with modest dust obscuration 
(A$_{\rm V}$ $<$ 20 mag) is $\sim$3.4 $\times$ 10$^{-3}$ \citep{pee04}.
Col.(9): The 11.3 $\mu$m PAH to infrared luminosity ratio in units of 
10$^{-3}$.
The ratio for normal starbursts with modest dust obscuration 
(A$_{\rm V}$ $<$ 20 mag) is $\sim$1.4 $\times$ 10$^{-3}$ \citep{soi02}.
}

\end{deluxetable}

\begin{deluxetable}{lccc}
\tabletypesize{\small}
\tablecaption{Optical depth of the 9.7 $\mu$m and 18 $\mu$m silicate
dust absorption feature \label{tbl-4}}
\tablewidth{0pt}
\tablehead{
\colhead{Object} & \colhead{$\tau_{9.7}'$} & \colhead{$\tau_{18}'$} & 
\colhead{$\tau_{18}'$/$\tau_{9.7}'$}  \\
\colhead{(1)} & \colhead{(2)} & \colhead{(3)} & \colhead{(4)} 
}
\startdata 
IRAS 04074$-$2801 & 3.0 & 1.2 & 0.40  \\ 
IRAS 05020$-$2941 & 2.4 & 0.9 & 0.38 \\
IRAS 13106$-$0922 & 2.0 & 1.2 & 0.60 \\ 
IRAS 14121$-$0126 & 1.3 & \nodata & \nodata \\
IRAS 21477+0502   & 0.8 & \nodata & \nodata \\ 
IRAS 03209$-$0806 & 1.0 & \nodata & \nodata \\
IRAS 10594+3818   & 1.0 & \nodata & \nodata \\
IRAS 12447+3721   & 1.4 & 0.5 & 0.36 \\
IRAS 14202+2615   & 0.7 & \nodata & \nodata \\
IRAS 15043+5754   & 1.4 & 0.7 & 0.50 \\
IRAS 22088$-$1831 & 2.6 & 1.0 & 0.38 \\
IRAS 02480$-$3745 & 1.4 & 0.6 & 0.43 \\
IRAS 08591+5248   & 1.0 & \nodata & \nodata \\
IRAS 02021$-$2103 & 1.4 & 0.4 & 0.29 \\
IRAS 08474+1813   & 2.0 & 1.0 & 0.50 \\
IRAS 14197+0813   & 0.8 & \nodata & \nodata  \\
IRAS 14485$-$2434 & 1.2 & \nodata & \nodata \\ \hline
\enddata

\tablecomments{
Col.(1): Object name.  
Col.(2): $\tau_{9.7}'$ is the optical depth of the 9.7 $\mu$m silicate
dust absorption feature, plotted against a power-law continuum, shown
as dotted lines in Figure 1.  
Col.(3): $\tau_{18}'$ is the optical depth of the 18 $\mu$m silicate
dust absorption feature, plotted against a power-law continuum, shown as
dotted lines in Figure 1. 
Col.(4): $\tau_{18}'$/$\tau_{9.7}'$ ratio for ULIRGs with clearly
detectable 18 $\mu$m silicate absorption.
The uncertainty with $\sim$10\% may be present.
}

\end{deluxetable}

\begin{deluxetable}{lcc}
\tablecaption{Ice absorption feature \label{tbl-5}}
\tablewidth{0pt}
\tablehead{
\colhead{Object} & \colhead{$\tau_{6.0}$} \\
\colhead{(1)} & \colhead{(2)}  
}
\startdata 
IRAS 04074$-$2801 & 0.3 \\ 
IRAS 05020$-$2941 & 0.4 \\
IRAS 13106$-$0922 & 1.0 \\ 
IRAS 14121$-$0126 & 0.8 \\
IRAS 21477+0502   & 0.6 \\ 
IRAS 22088$-$1831 & 0.5 \\
IRAS 02480$-$3745 & 0.9 \\
IRAS 08591+5248   & 0.5 \\
IRAS 14485$-$2434 & 0.3 \\ \hline
\enddata

\tablecomments{
Col.(1): Object name.  
Col.(2): Optical depth of the 6.0 $\mu$m H$_{2}$O ice absorption feature for
clearly detected sources. 
The uncertainty with $\sim$0.1 may be present. 
}

\end{deluxetable}

\begin{deluxetable}{lcccc}
\tabletypesize{\scriptsize}
\tablecaption{Buried AGN signatures \label{tbl-6}}
\tablewidth{0pt}
\tablehead{
\colhead{Object} & \colhead{EW$_{\rm 6.2PAH}$} & 
\colhead{EW$_{\rm 11.3PAH}$} & \colhead{$\tau_{9.7}'$} & 
\colhead{Total} \\
\colhead{(1)} & \colhead{(2)} & \colhead{(3)} &  
\colhead{(4)} & \colhead{(5)}\\  
}
\startdata 
IRAS 04074$-$2801 & $\bigcirc$ & X & $\bigcirc$ & $\bigcirc$ \\ 
IRAS 05020$-$2941 & $\bigcirc$ & X & $\bigcirc$ & $\bigcirc$ \\
IRAS 13106$-$0922 & $\bigcirc$ & X & $\bigcirc$ & $\bigcirc$ \\ 
IRAS 14121$-$0126 & X & X & X & X \\
IRAS 21477+0502   & X & $\bigcirc$ & X & $\bigcirc$ \\ 
IRAS 03209$-$0806 & X & X & X & X \\
IRAS 10594+3818   & X & X & X & X \\
IRAS 12447+3721   & X & X & X & X \\
IRAS 14202+2615   & $\bigcirc$ & X & X & $\bigcirc$ \\
IRAS 15043+5754   & X & X & X & X \\
IRAS 22088$-$1831 & $\bigcirc$ & $\bigcirc$ & $\bigcirc$ & $\bigcirc$ \\
IRAS 02480$-$3745 & X & X & X & X \\
IRAS 08591+5248   & X & X & X & X \\
IRAS 02021$-$2103 & X & X & X & X \\
IRAS 08474+1813   & $\bigcirc$ & X & $\bigcirc$ & $\bigcirc$ \\
IRAS 14197+0813   & X & X & X & X \\
IRAS 14485$-$2434 & $\bigcirc$ & X & X & $\bigcirc$ \\ \hline
\enddata

\tablecomments{
Col.(1): Object name.  
Col.(2): Buried AGN signatures based on the low equivalent width of the
         6.2 $\mu$m PAH emission (EW$_{\rm 6.2PAH}$ $<$ 180 nm) ($\S$5.2.1).
         $\bigcirc$: present.  X: none. 
Col.(3): Buried AGN signatures based on the low equivalent width of the
         11.3 $\mu$m PAH emission (EW$_{\rm 11.3PAH}$ $<$ 200 nm) ($\S$5.2.1).
         $\bigcirc$: present.  X: none. 
Col.(4): Buried AGN signatures based on the large $\tau_{9.7}'$ value 
         ($\geq$2) ($\S$5.2.2).
         $\bigcirc$: present.  X: none. 
Col.(5): Buried AGN signatures from combined methods in Cols. (2)--(4). 
         $\bigcirc$: strong. X: none.
}

\end{deluxetable}

\begin{deluxetable}{lcccc}
\tablecaption{Luminosities of buried AGNs after extinction-correction 
and modestly-obscured starbursts \label{tbl-7}}
\tablewidth{0pt}
\tablehead{
\colhead{Object} & \colhead{L(AGN)} 
& \colhead{L(SB-6.2PAH)} & \colhead{L(SB-11.3PAH)} 
& \colhead{L$_{\rm IR}$} \\
\colhead{} & \colhead{10$^{45}$ [ergs s$^{-1}$]} & 
\colhead{10$^{45}$ [ergs s$^{-1}$]} & 
\colhead{10$^{45}$ [ergs s$^{-1}$]} & 
\colhead{10$^{45}$ [ergs s$^{-1}$]} \\
\colhead{(1)} & \colhead{(2)} & \colhead{(3)} & \colhead{(4)} & \colhead{(5)} 
}
\startdata 
IRAS 04074$-$2801 & 2   & 1   & 2   & 6 \\
IRAS 05020$-$2941 & 1.5 & 2   & 3   & 8 \\
IRAS 13106$-$0922 & 1   & 1   & 2.5 & 8 \\
IRAS 14202+2615   & 1   & 5.5 & 9.5 & 10 \\
IRAS 22088$-$1831 & 1.5 & 1   & 1.5 & 10 \\
IRAS 08474+1813   & 0.3 & 0.7 & 1   & 5 \\
IRAS 14485$-$2434 & 1   & 2   & 5   & 5 \\ \hline
\enddata

\tablecomments{
Col.(1): Object name.  
Col.(2): Extinction-corrected intrinsic luminosity of a buried AGN 
in units of 10$^{45}$ [ergs s$^{-1}$].
Col.(3): Infrared (8--1000 $\mu$m) luminosity of modestly-obscured
(A$_{\rm V}$ $<$ 20 mag) starbursts, estimated from the 6.2 $\mu$m PAH
emission luminosity (L$_{\rm 6.2PAH}$) and L$_{\rm 6.2PAH}$/L$_{\rm IR}$
= 3.4 $\times$ 10$^{-3}$ \citep{pee04}, in units of 10$^{45}$ 
[ergs s$^{-1}$].   
Col.(4): Infrared (8--1000 $\mu$m) luminosity of modestly-obscured
(A$_{\rm V}$ $<$ 20 mag) starbursts, estimated from the 11.3 $\mu$m PAH
emission luminosity (L$_{\rm 11.3PAH}$) and L$_{\rm 11.3PAH}$/L$_{\rm
IR}$ = 1.4 $\times$ 10$^{-3}$ \citep{soi02}, in units of 10$^{45}$ 
[ergs s$^{-1}$].  
Col.(5): Observed infrared luminosity in units of
10$^{45}$ [ergs s$^{-1}$]. 
}

\end{deluxetable}

\begin{deluxetable}{llcc}
\tablecaption{Buried AGN fraction in ULIRGs \label{tbl-8}}
\tablewidth{0pt}
\tablehead{
\colhead{Optical classification} & \colhead{Sub-category} 
& \colhead{number of sources} & \colhead{Buried AGNs} \\
\colhead{(1)} & \colhead{(2)} & \colhead{(3)} & \colhead{(4)} 
} 
\startdata 
non-Seyfert  & total & 85 & 37 (44\%) \\
             & L$_{\rm IR}$ $<$ 10$^{12.3}$L$_{\odot}$    & 54 & 15 (28\%) \\
             & L$_{\rm IR}$ $\geq$ 10$^{12.3}$L$_{\odot}$ & 31 & 22 (71\%) \\
             & $z \leq$ 0.15 & 52 & 18 (35\%) \\
             & $z >$ 0.15 & 33 & 19 (58\%) \\ \hline
LINER        & total & 43 & 21 (49\%) \\
             & L$_{\rm IR}$ $<$ 10$^{12.3}$L$_{\odot}$ & 25 & 9 (36\%) \\
             & L$_{\rm IR}$ $\geq$ 10$^{12.3}$L$_{\odot}$ & 18 & 12 (67\%) \\
             & $z \leq$ 0.15 & 28 & 10 (36\%) \\
             & $z >$ 0.15 & 15 & 11 (73\%) \\ \hline
HII-region   & total & 32 & 12 (38\%) \\
             & L$_{\rm IR}$ $<$ 10$^{12.3}$L$_{\odot}$ & 22 &  4 (18\%) \\
             & L$_{\rm IR}$ $\geq$ 10$^{12.3}$L$_{\odot}$  & 10 & 8 (80\%) \\
             & $z \leq$ 0.15 & 20 & 6 (30\%) \\
             & $z >$ 0.15 & 12 & 6 (50\%) \\ \hline
Unclassified & total & 10 & 4 (40\%) \\
             & L$_{\rm IR}$ $<$ 10$^{12.3}$L$_{\odot}$ & 7 & 2 (29\%) \\
             & L$_{\rm IR}$ $\geq$ 10$^{12.3}$L$_{\odot}$ & 3 & 2 (67\%) \\
             & $z \leq$ 0.15 & 4 & 2 (50\%) \\
             & $z >$ 0.15 & 6 & 2 (33\%) \\ \hline
\enddata

\tablecomments{
Col.(1): Optical classification.  
Col.(2): Sub-category of ULIRGs. Whether 
10$^{12}$L$_{\odot}$ $\leq$ L$_{\rm IR}$ $<$ 10$^{12.3}$L$_{\odot}$ 
or L$_{\rm IR}$ $\geq$ 10$^{12.3}$L$_{\odot}$, and whether 
$z \leq$ 0.15 or $z >$ 0.15.
Col.(3): Number of sources in each sub-category.
Col.(4): Number and fraction of ULIRGs with strong buried AGN signatures
in each sub-category (Imanishi et al. 2007; Imanishi 2009, this paper).
}

\end{deluxetable}

\clearpage

\begin{figure}
\includegraphics[angle=-90,scale=.35]{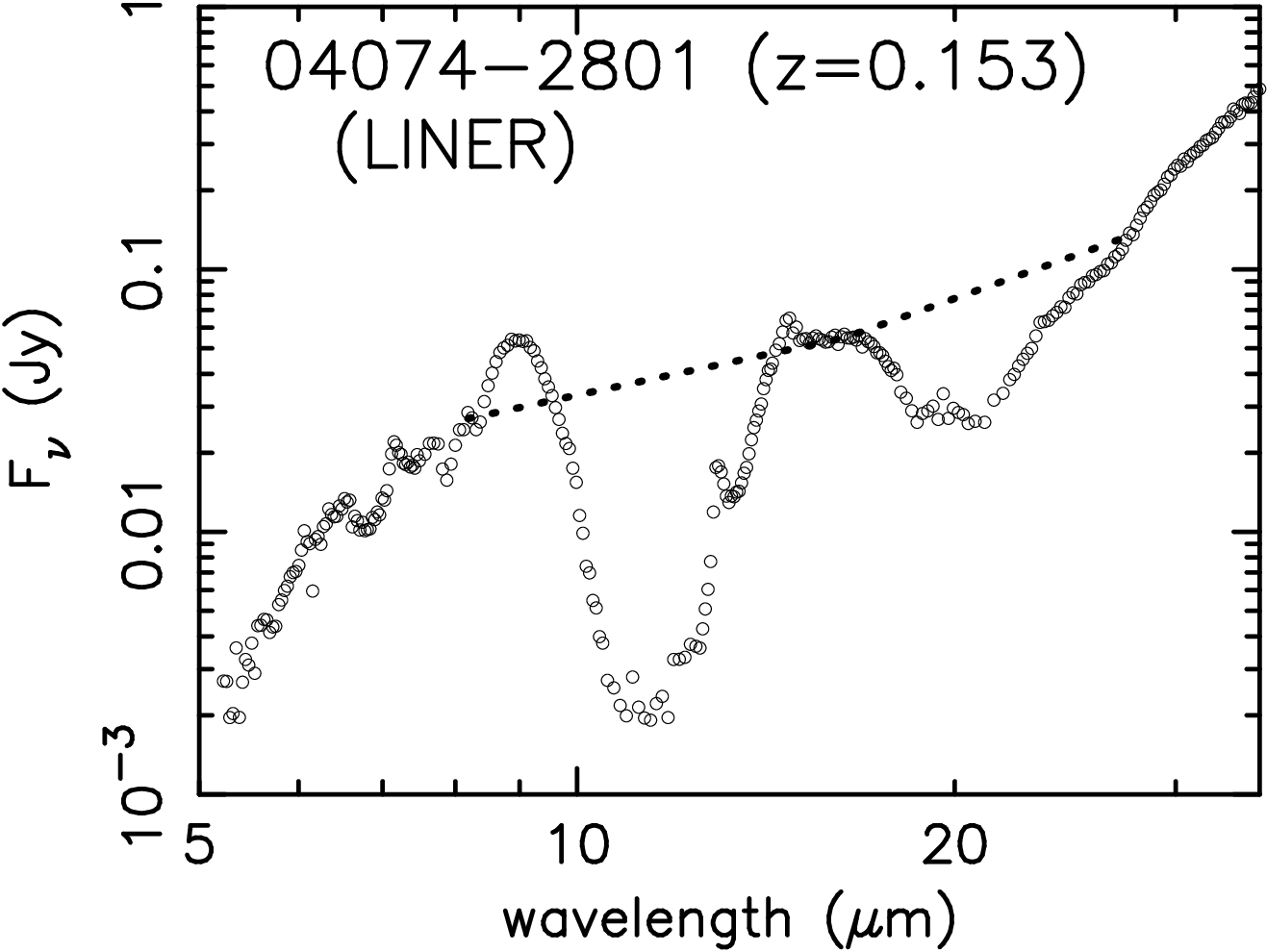} \hspace{0.3cm}
\includegraphics[angle=-90,scale=.35]{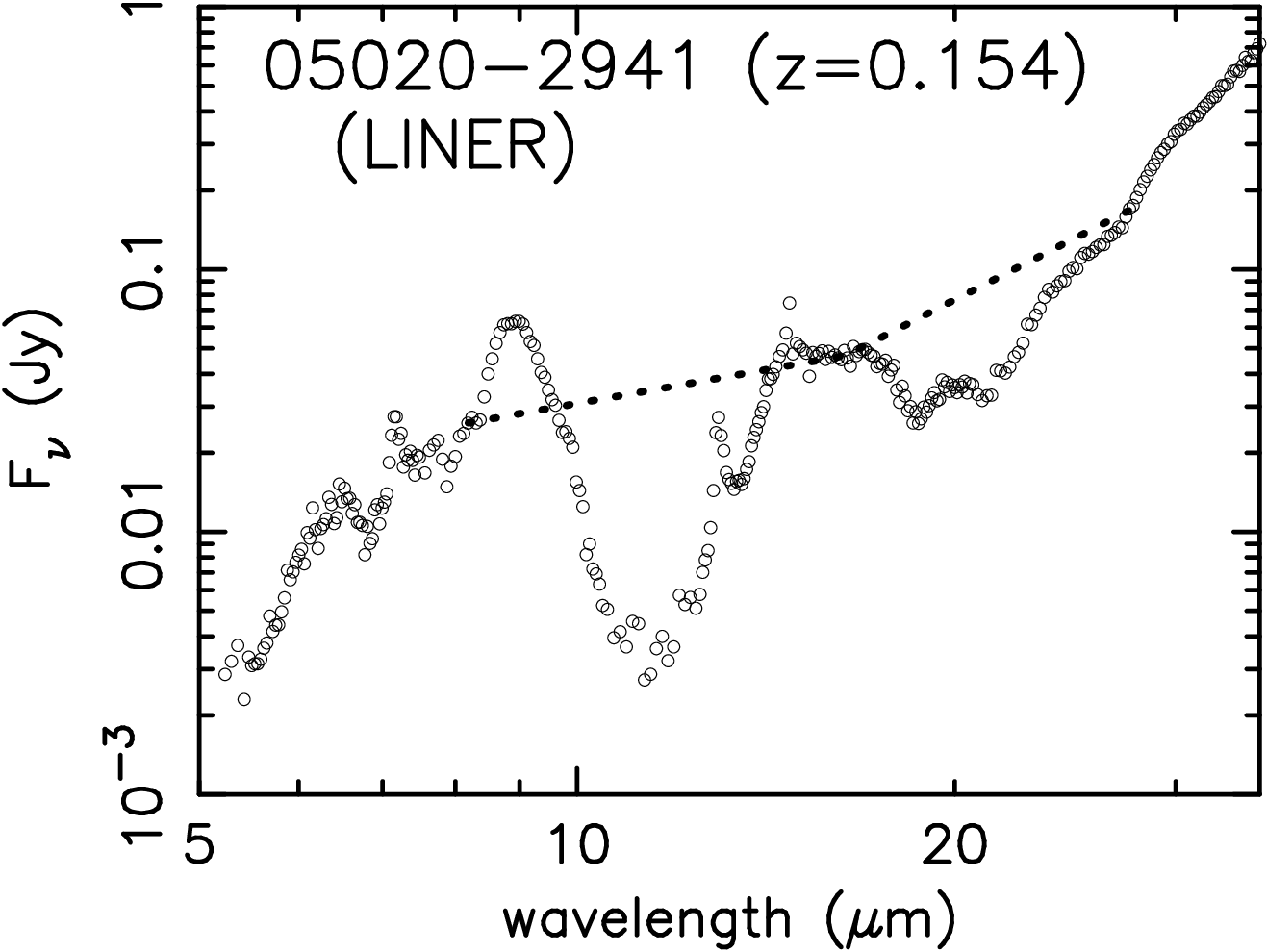} \\
\includegraphics[angle=-90,scale=.35]{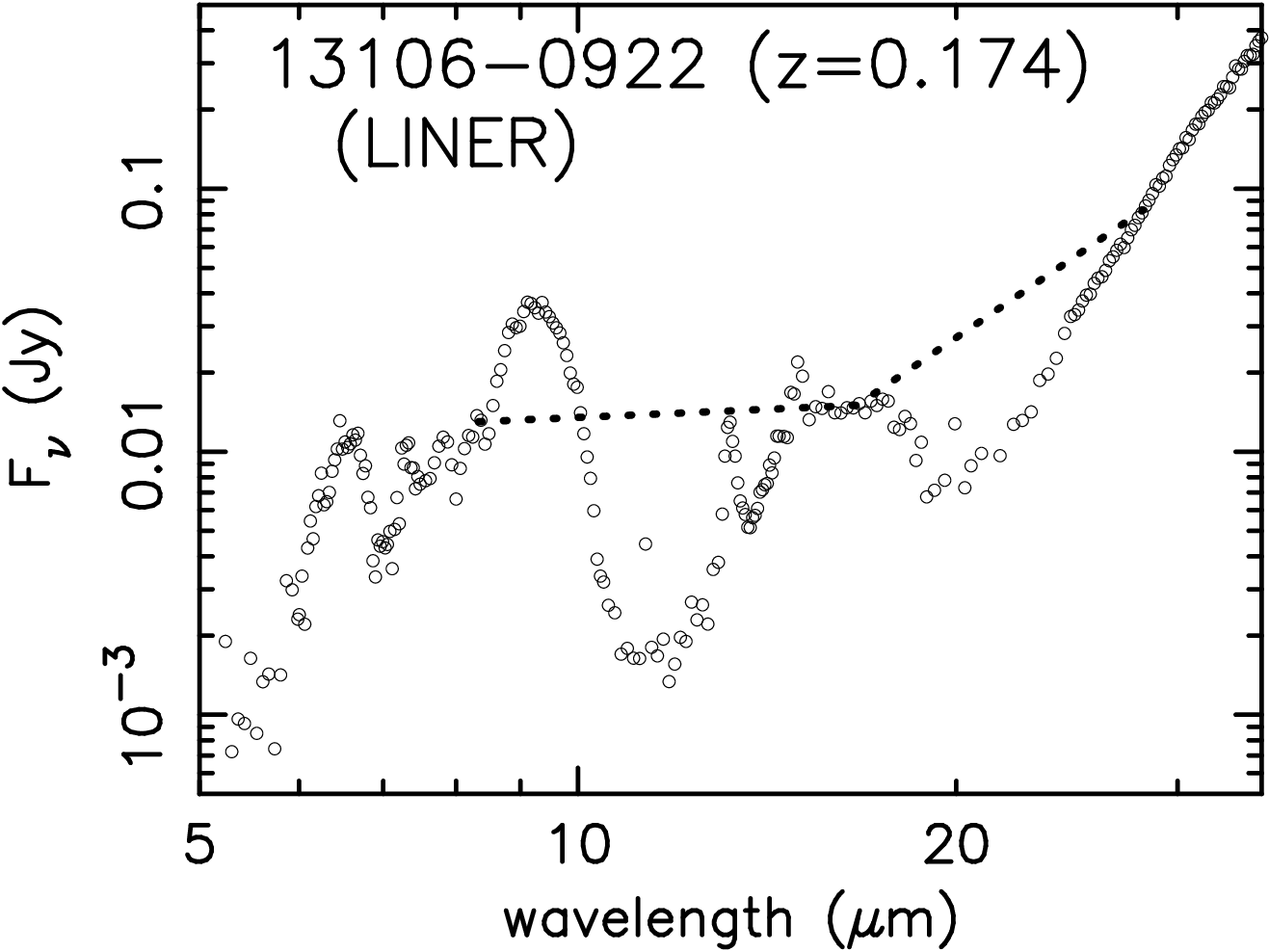} \hspace{0.3cm}
\includegraphics[angle=-90,scale=.35]{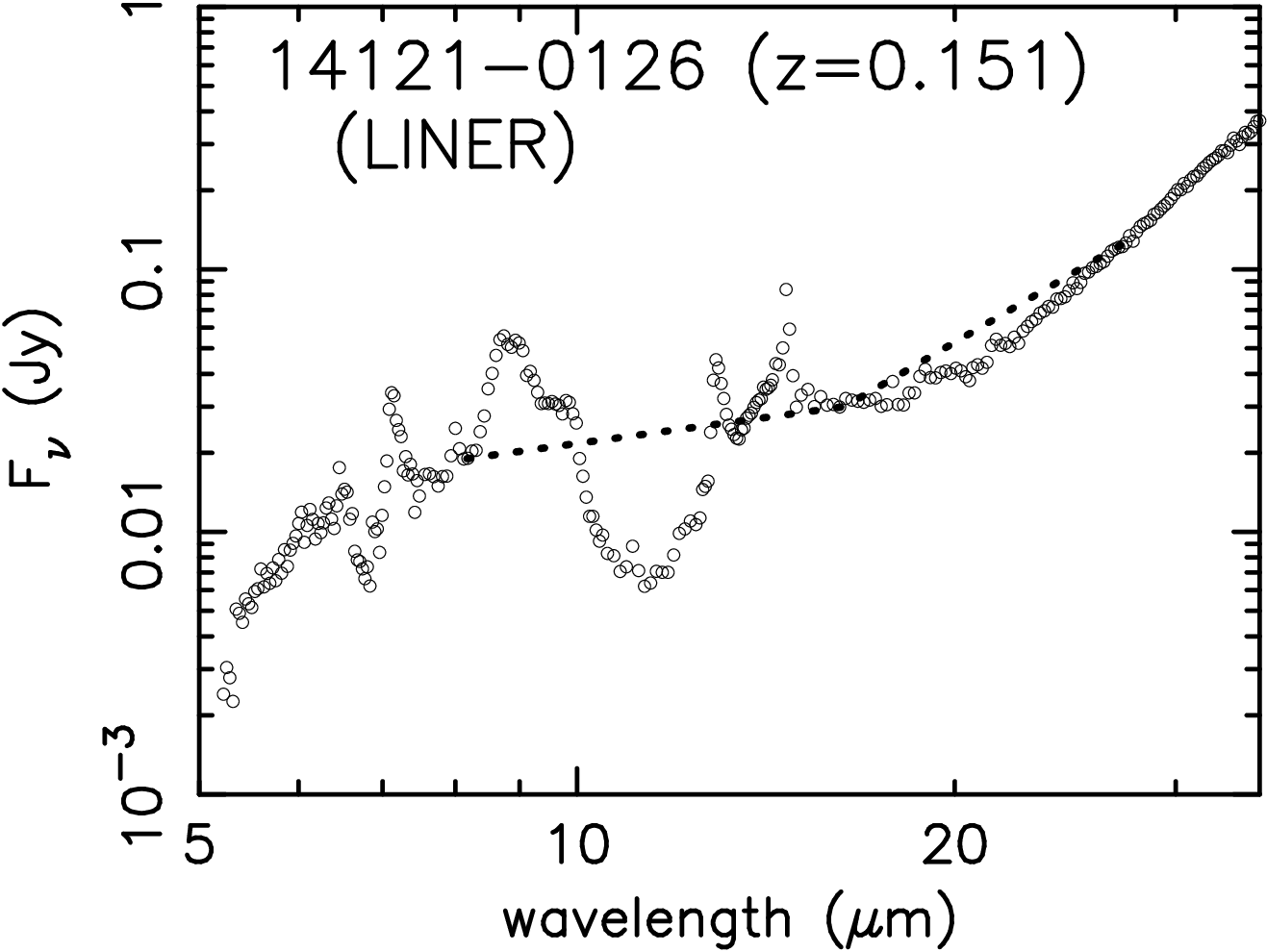} \\
\includegraphics[angle=-90,scale=.35]{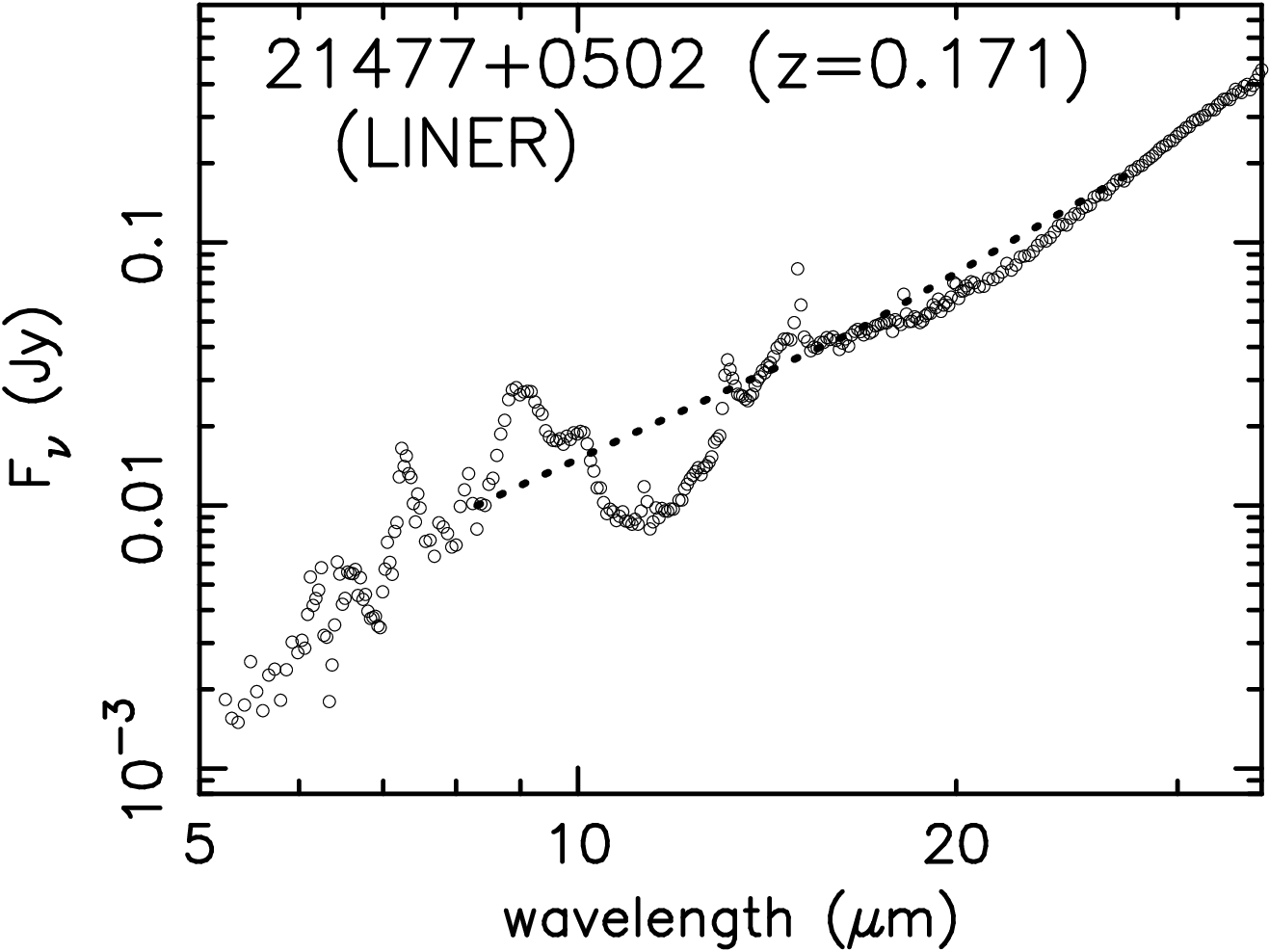} \hspace{0.3cm} 
\end{figure}

\clearpage

\begin{figure}
\includegraphics[angle=-90,scale=.35]{f1f.eps} \hspace{0.3cm}
\includegraphics[angle=-90,scale=.35]{f1g.eps} \\
\includegraphics[angle=-90,scale=.35]{f1h.eps} \hspace{0.3cm}
\includegraphics[angle=-90,scale=.35]{f1i.eps} \\
\includegraphics[angle=-90,scale=.35]{f1j.eps} \hspace{0.3cm} 
\includegraphics[angle=-90,scale=.35]{f1k.eps} \\
\end{figure}

\clearpage

\begin{figure}
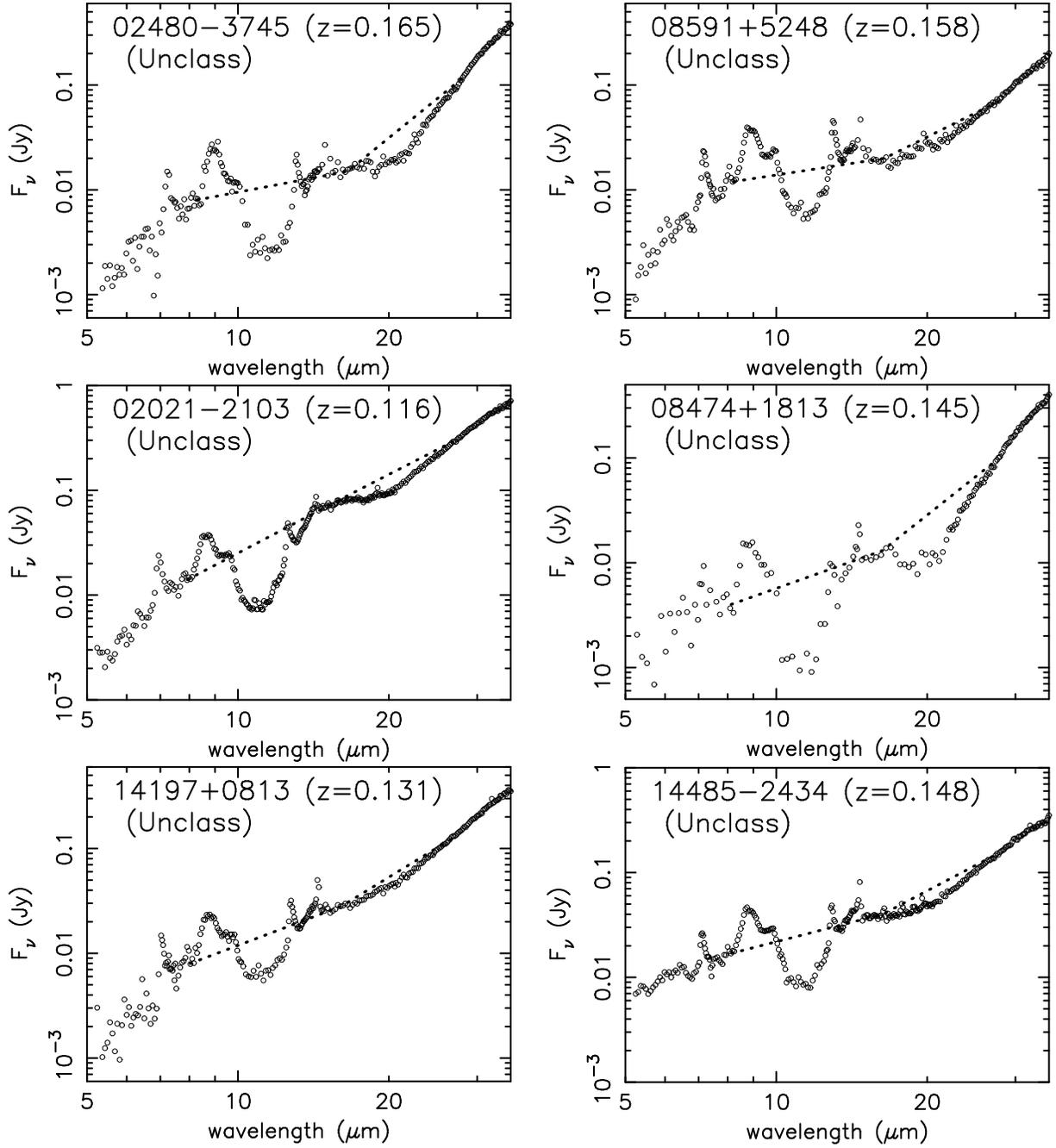

\includegraphics[angle=-90,scale=.35]{f1l.eps} \hspace{0.3cm}
\includegraphics[angle=-90,scale=.35]{f1m.eps} \\
\includegraphics[angle=-90,scale=.35]{f1n.eps} \hspace{0.3cm}
\includegraphics[angle=-90,scale=.35]{f1o.eps} \\
\includegraphics[angle=-90,scale=.35]{f1p.eps} \hspace{0.3cm}
\includegraphics[angle=-90,scale=.35]{f1q.eps} \\
\caption{
Infrared 5--35 $\mu$m spectra of optically non-Seyfert ULIRGs, taken
with {\it Spitzer} IRS. 
The abscissa and ordinate are, respectively, the observed wavelength in
$\mu$m and the flux F$_{\nu}$ in Jy, both plotted in a decimal
logarithmic scale.   
For all objects, the ratio of the uppermost to the lowermost scale in the
ordinate is fixed as a factor of 1000, to illustrate the variation of
the overall spectral energy distribution.
Dotted line: power-law continuum determined from data points at
$\lambda_{\rm rest}$ = 7.1 $\mu$m and 14.2 $\mu$m (in the rest frame)  
for the 9.7 $\mu$m silicate dust absorption feature, and at
$\lambda_{\rm rest}$ = 14.2 $\mu$m and 24 $\mu$m for the 18 $\mu$m 
silicate dust absorption feature (see $\S$4.2).
}
\end{figure}

\begin{figure}
\includegraphics[angle=-90,scale=.35]{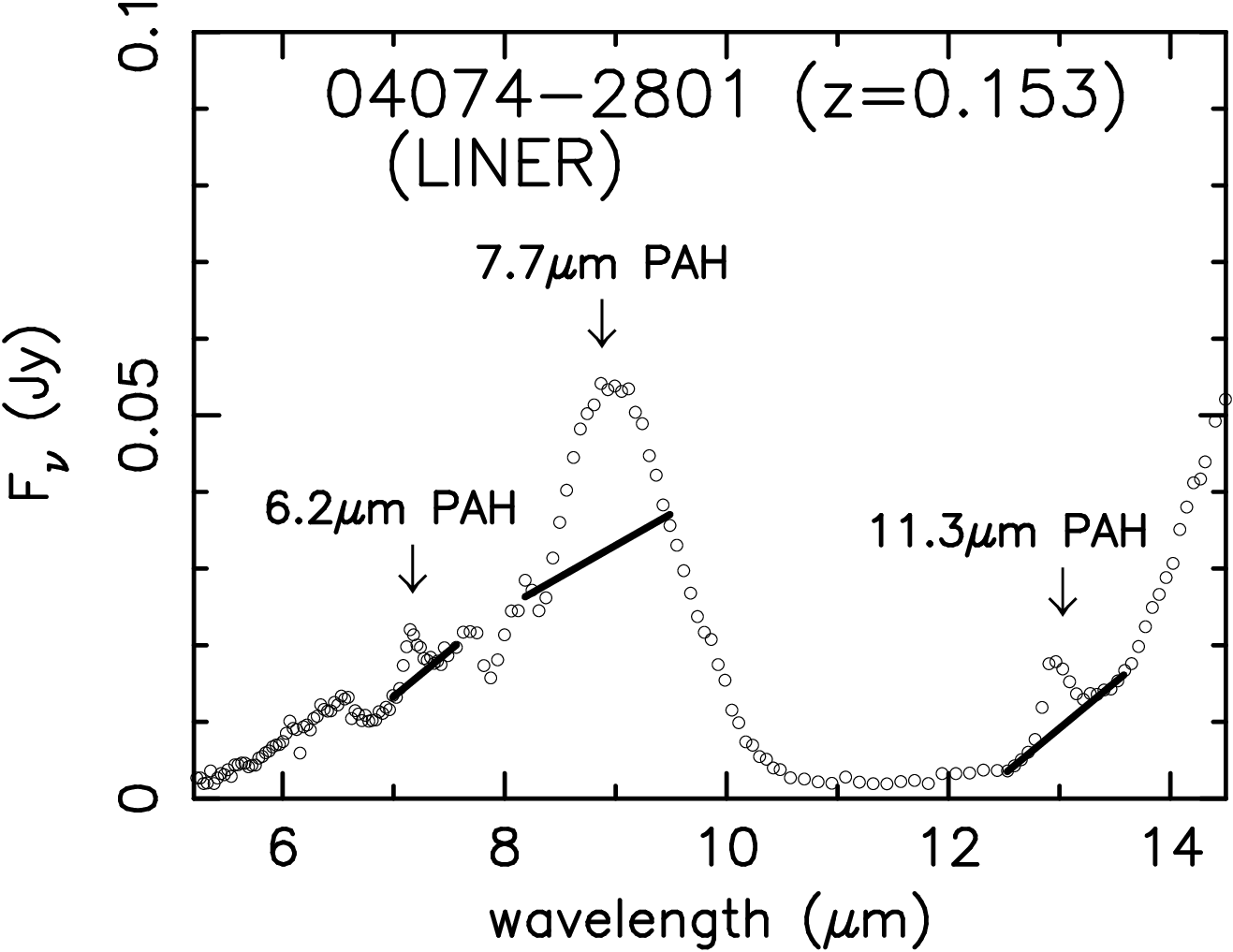} \hspace{0.3cm}
\includegraphics[angle=-90,scale=.35]{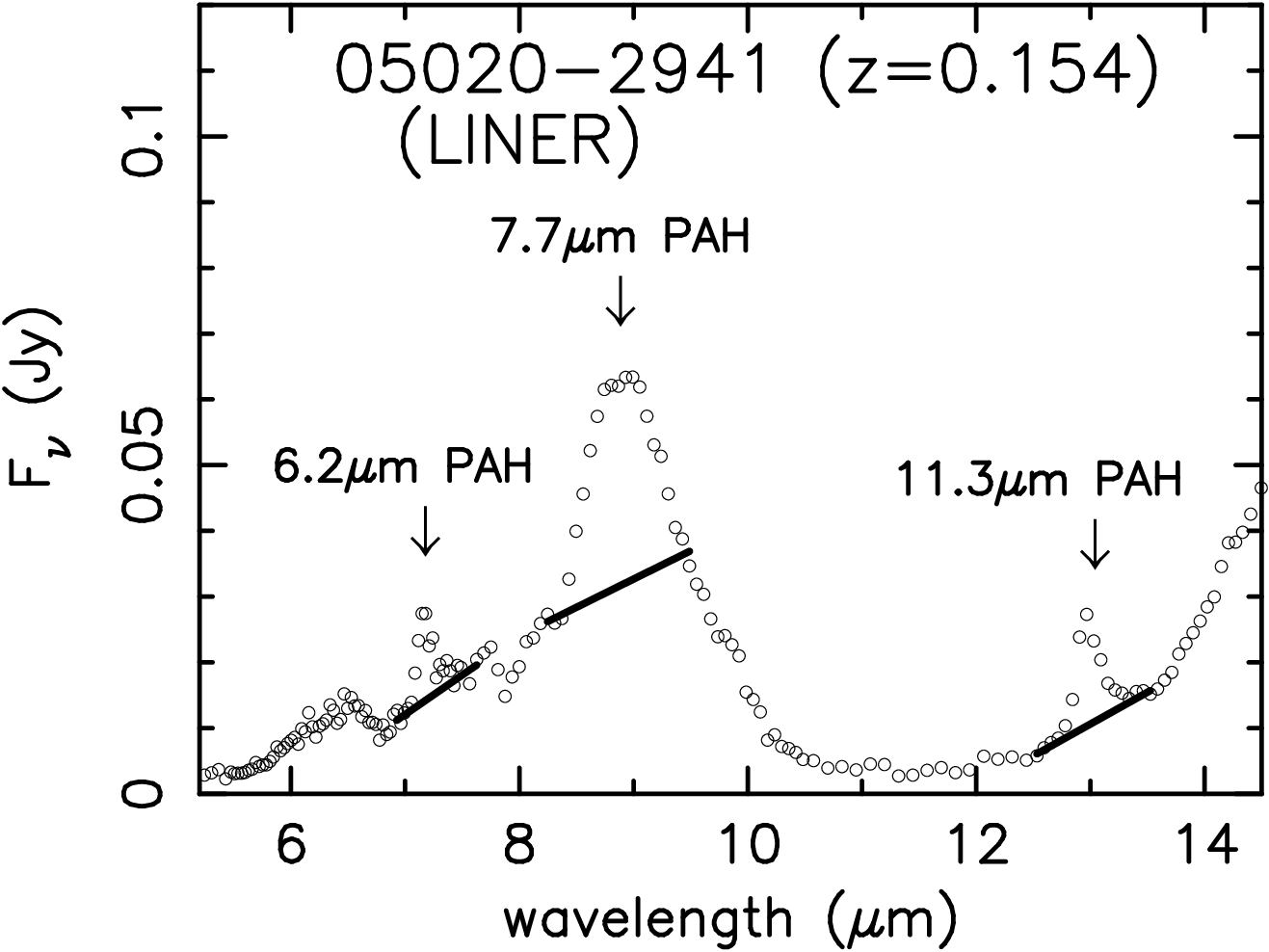} \\
\includegraphics[angle=-90,scale=.35]{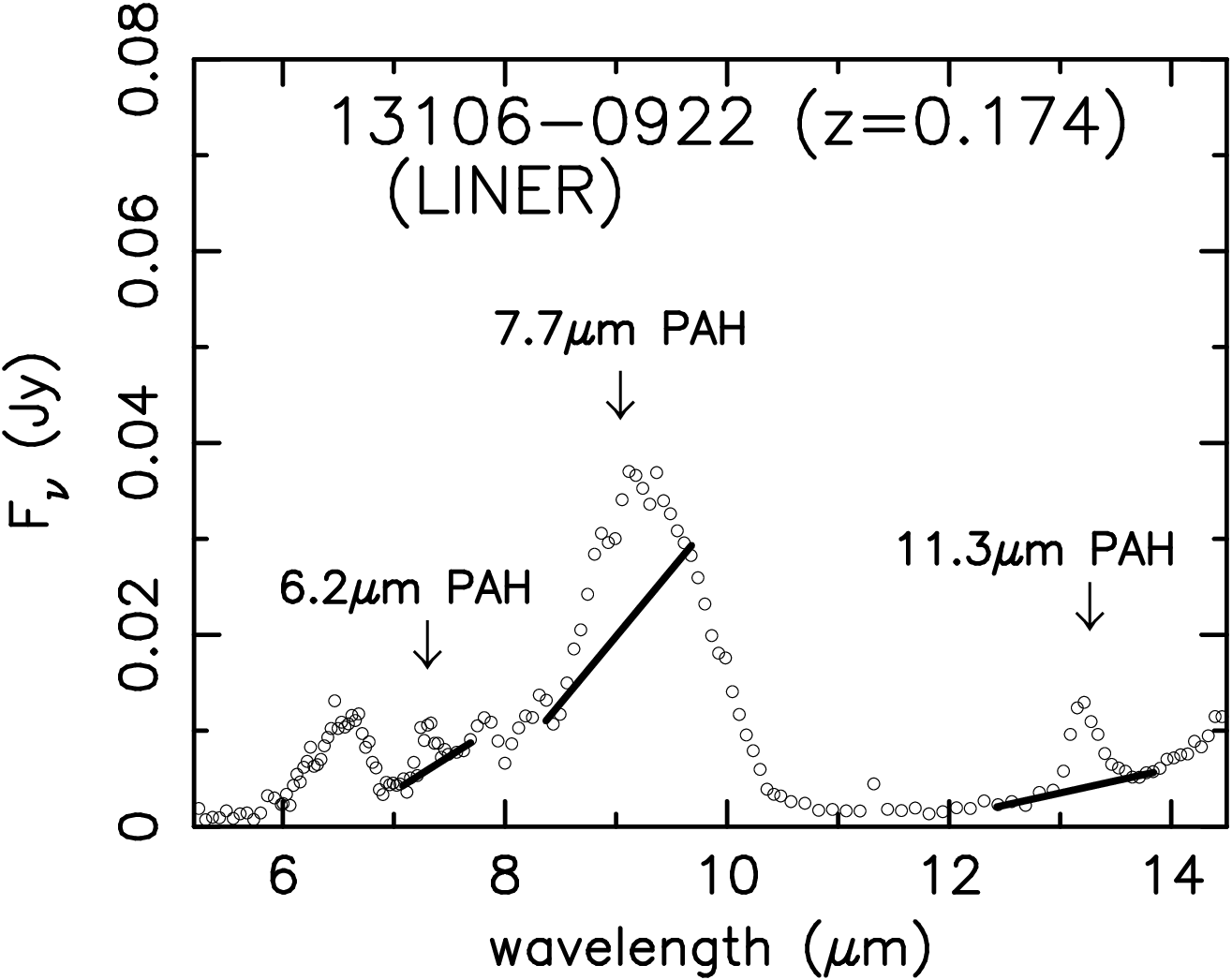} \hspace{0.3cm}
\includegraphics[angle=-90,scale=.35]{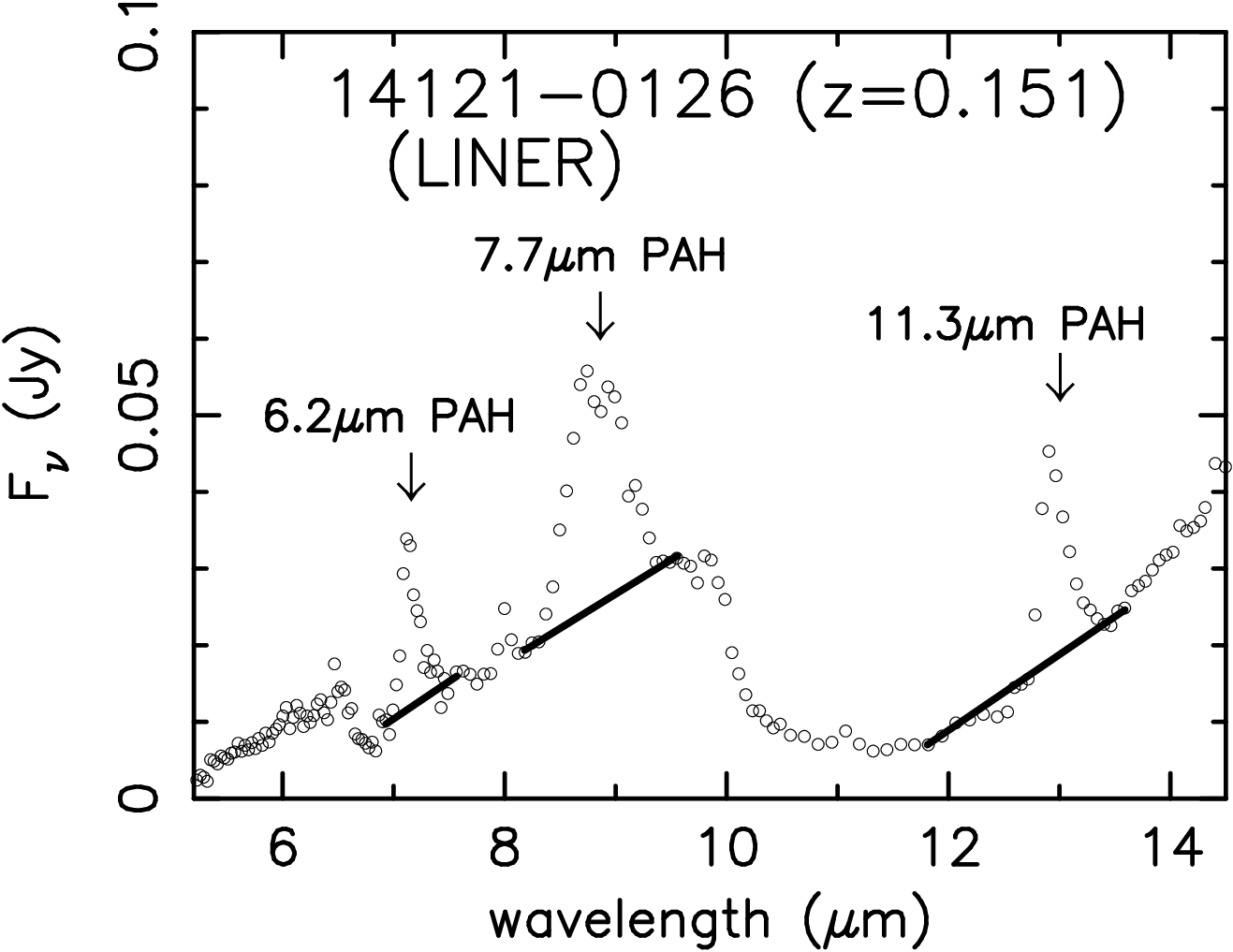} \\
\includegraphics[angle=-90,scale=.35]{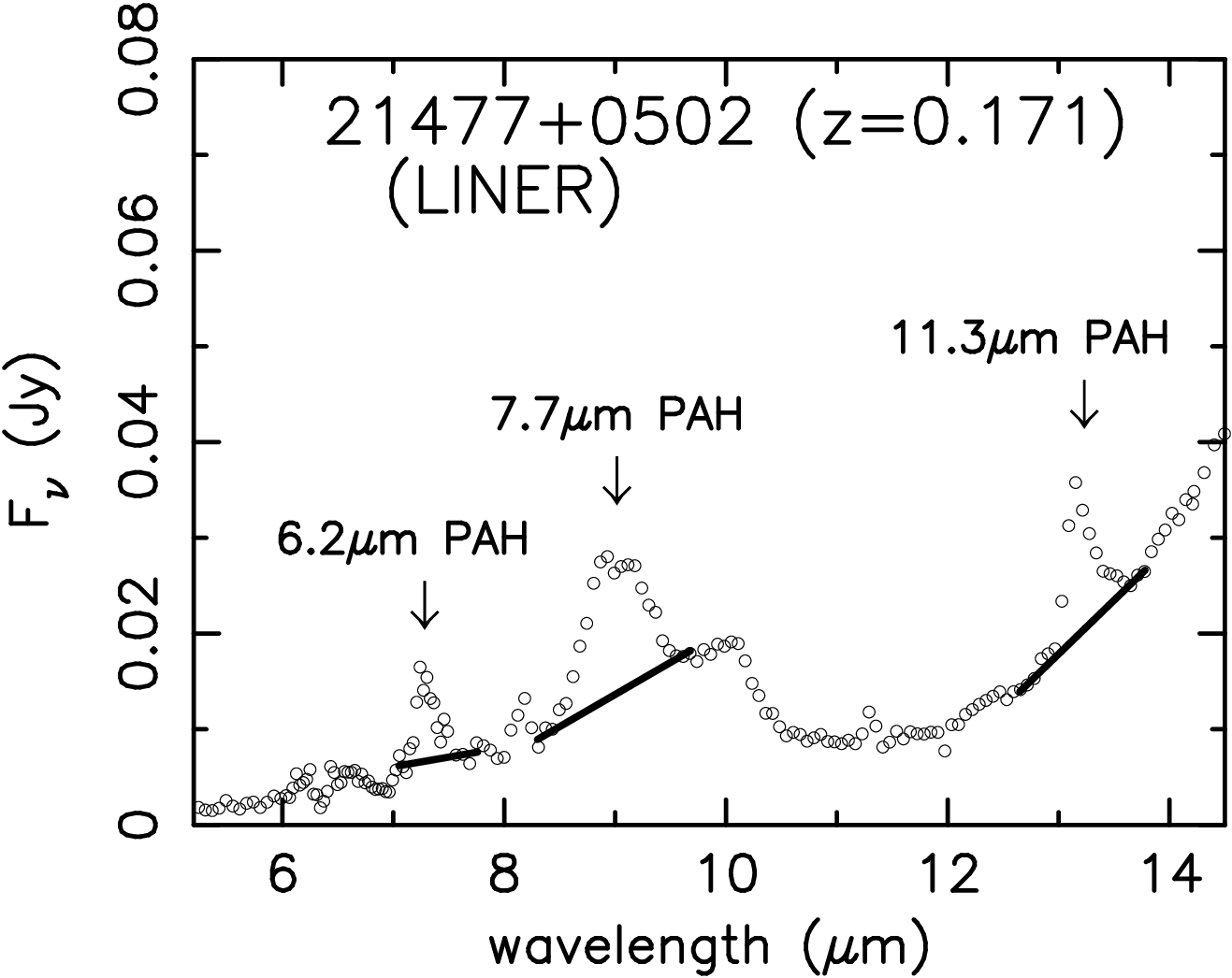} \hspace{0.3cm} 
\end{figure}

\clearpage

\begin{figure}
\includegraphics[angle=-90,scale=.35]{f2f.eps} \hspace{0.3cm}
\includegraphics[angle=-90,scale=.35]{f2g.eps} \\
\includegraphics[angle=-90,scale=.35]{f2h.eps} \hspace{0.3cm}
\includegraphics[angle=-90,scale=.35]{f2i.eps} \\
\includegraphics[angle=-90,scale=.35]{f2j.eps} \hspace{0.3cm} 
\includegraphics[angle=-90,scale=.35]{f2k.eps} \\
\end{figure}

\clearpage

\begin{figure}
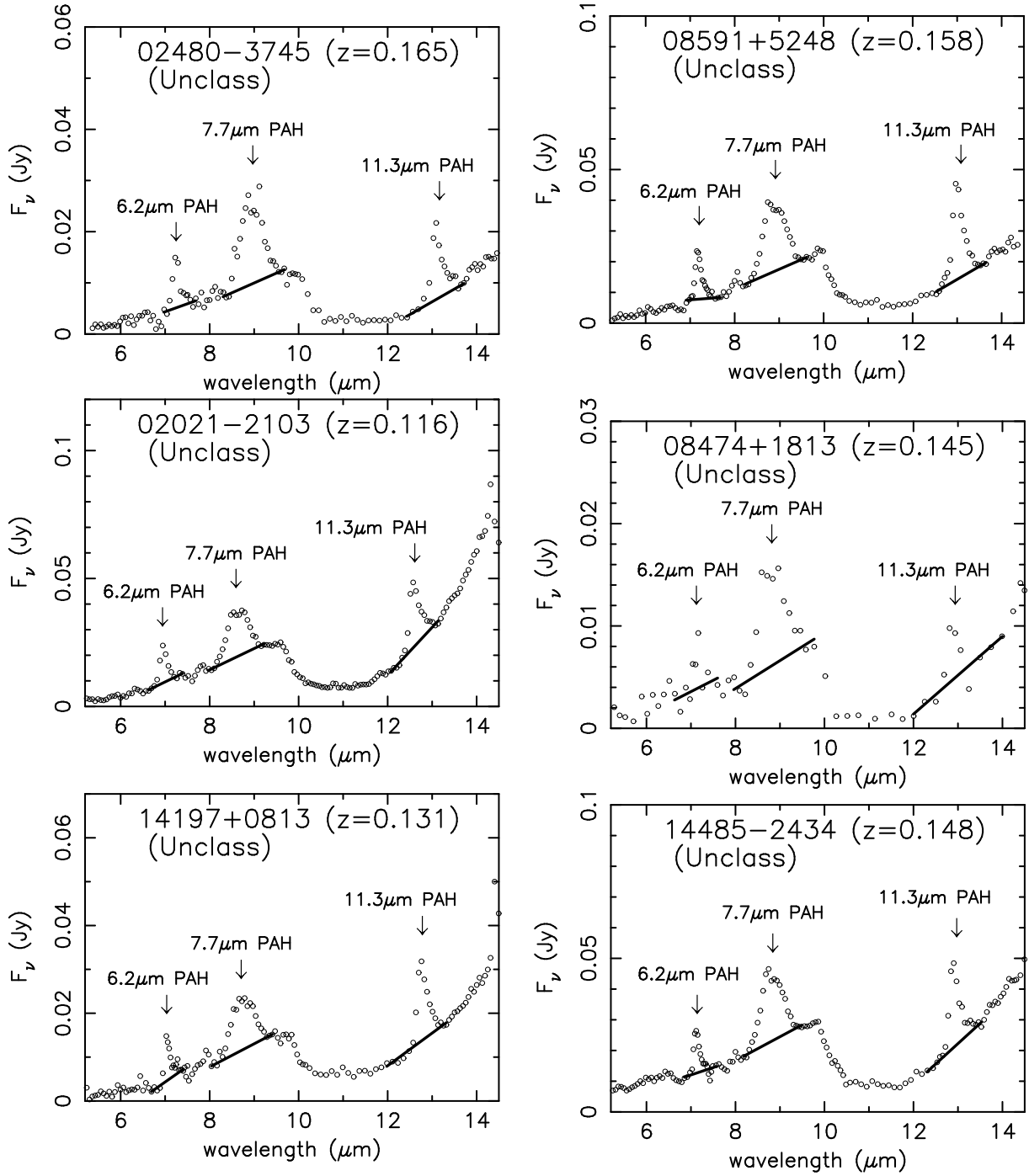

\includegraphics[angle=-90,scale=.35]{f2l.eps} \hspace{0.3cm}
\includegraphics[angle=-90,scale=.35]{f2m.eps} \\
\includegraphics[angle=-90,scale=.35]{f2n.eps} \hspace{0.3cm}
\includegraphics[angle=-90,scale=.35]{f2o.eps} \\
\includegraphics[angle=-90,scale=.35]{f2p.eps} \hspace{0.3cm}
\includegraphics[angle=-90,scale=.35]{f2q.eps} \\
\caption{
{\it Spitzer} IRS spectra of ULIRGs at $\lambda_{\rm obs}$ = 5.2--14.5
$\mu$m, for investigating the PAH emission in detail.  The
abscissa and ordinate are, respectively, the observed 
wavelength in $\mu$m and the flux in Jy, both plotted in a linear scale.
The expected wavelengths of the 6.2 $\mu$m, 7.7 $\mu$m, and 11.3 $\mu$m
PAH emission features are indicated as down arrows with labels.  
The solid lines are the adopted linear continuum levels for estimating the
strength of the PAH emission (see $\S$4.1).
}
\end{figure}

\begin{figure}
\includegraphics[angle=-90,scale=.35]{f3a.eps} \hspace{0.3cm} 
\includegraphics[angle=-90,scale=.35]{f3b.eps} \\
\includegraphics[angle=-90,scale=.35]{f3c.eps} \hspace{0.3cm} 
\includegraphics[angle=-90,scale=.35]{f3d.eps} \\
\includegraphics[angle=-90,scale=.35]{f3e.eps} \hspace{0.3cm} 
\includegraphics[angle=-90,scale=.35]{f3f.eps} \\ 
\end{figure}

\begin{figure}
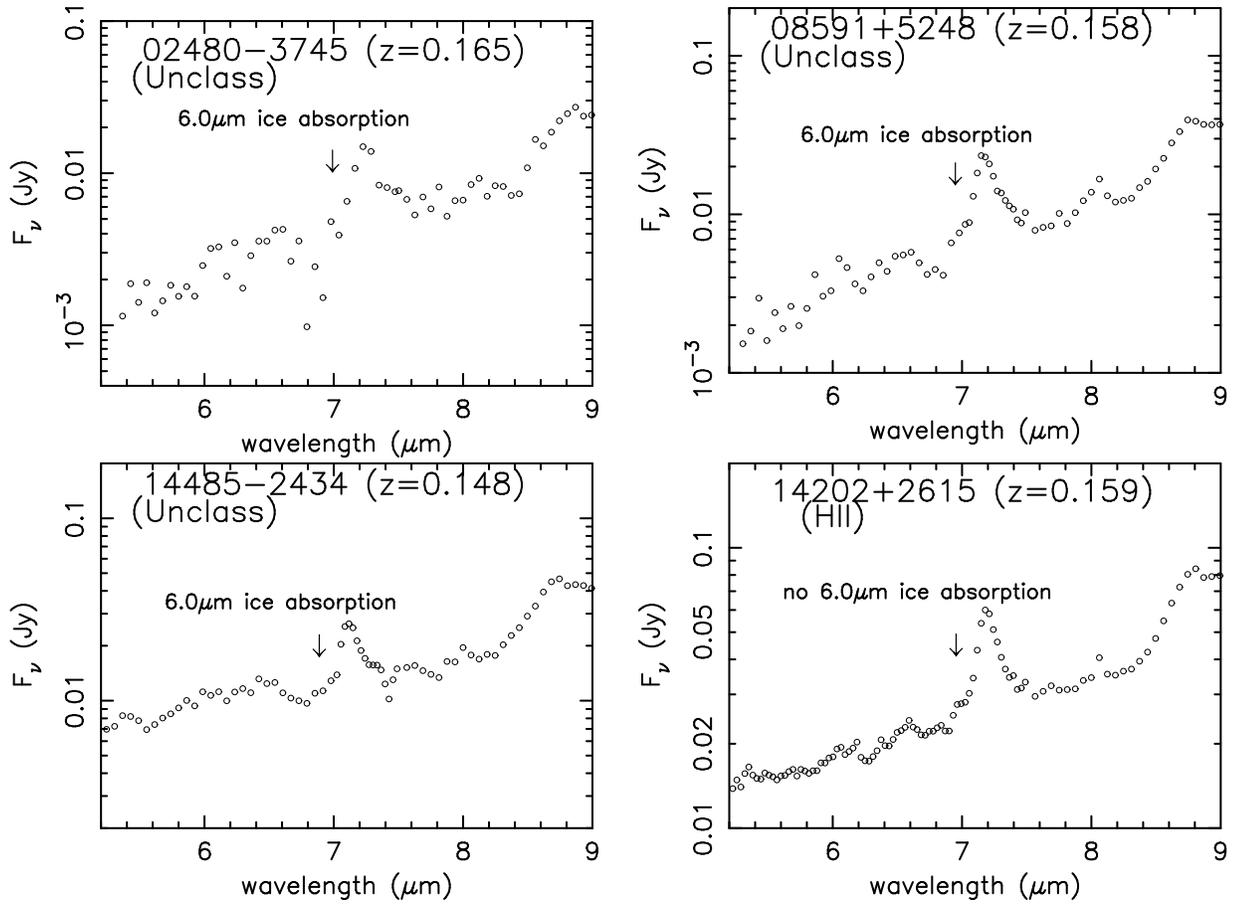

\includegraphics[angle=-90,scale=.35]{f3g.eps} \hspace{0.3cm}
\includegraphics[angle=-90,scale=.35]{f3h.eps} \\ 
\includegraphics[angle=-90,scale=.35]{f3i.eps} \hspace{0.3cm}
\includegraphics[angle=-90,scale=.35]{f3j.eps} \\ 
\caption{
{\it Spitzer} IRS spectra at $\lambda_{\rm obs}$ = 5.2--9 $\mu$m 
for ULIRGs
displaying clear 6.0 $\mu$m H$_{2}$O ice absorption features (marked with
``6.0$\mu$m ice absorption'' in the first nine plots).  
The spectrum of IRAS 14202+2615, marked ``no 6.0$\mu$m ice
absorption'', is shown as an example of undetected ice absorption.  
The abscissa is the observed wavelength in $\mu$m plotted in a linear
scale, and the ordinate is the flux in Jy plotted in a decimal
logarithmic scale.} 
\end{figure}

\clearpage 

\begin{figure}
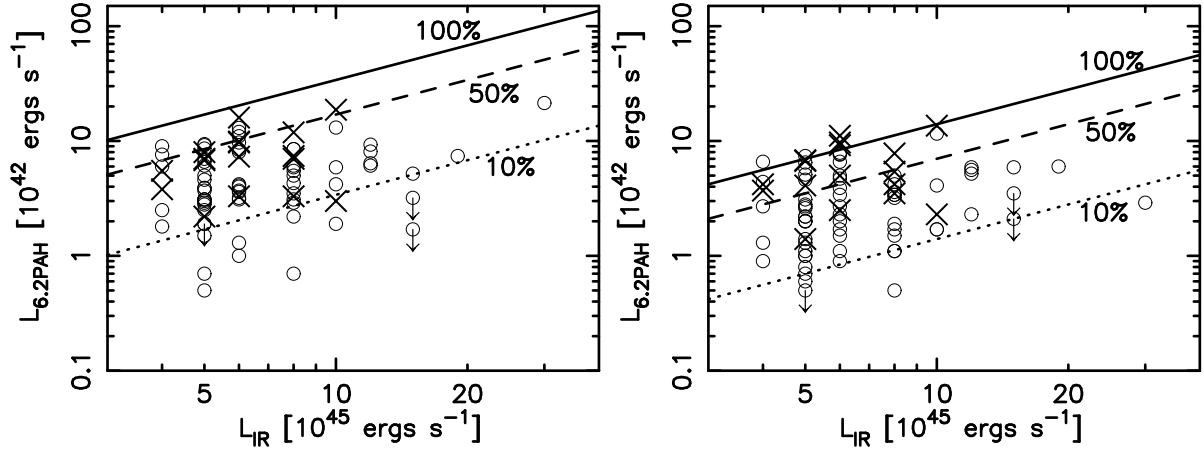

\includegraphics[angle=-90,scale=.35]{f4a.eps} 
\includegraphics[angle=-90,scale=.35]{f4b.eps} 
\caption{
Comparison of the 6.2 $\mu$m (Left) and 11.3 $\mu$m (Right) PAH
luminosities with the infrared luminosities. 
Optically non-Seyfert ULIRGs that were newly studied in this paper are plotted
with ``X'' symbols.  
Open circles represent sources studied in previously published papers 
\citep{ima07a,ima09}.
The solid lines indicate the canonical PAH to infrared luminosity ratios
found in modestly obscured starburst galaxies (see $\S$5.1). 
Specifically, on the solid lines, 100\% of the infrared luminosity can be
reproduced from the detected modestly obscured starburst activity. 
The dashed and dotted lines represent 50\% and 10\% of the ratios.}
\end{figure}

\begin{figure}
\includegraphics[angle=-90,scale=.35]{f5a.eps}  \hspace{0.3cm}
\includegraphics[angle=-90,scale=.35]{f5b.eps} \\
\includegraphics[angle=-90,scale=.35]{f5c.eps} 
\caption{
Distribution of {\it (a)} EW$_{\rm 6.2PAH}$, 
{\it (b)} EW$_{\rm 11.3PAH}$, and {\it (c)} $\tau_{9.7}'$, as a function
of galaxy infrared luminosity. 
Optically non-Seyfert ULIRGs that were newly studied in this paper are plotted
with ``X'' symbols.  
Open circles represent sources studied in previously published papers 
\citep{ima07a,ima09,bra06}.
The horizontal dashed lines indicate the threshold for buried AGN
candidates ($\S$5.2.1 and 5.2.2).
For the plots (a) and (b), \citet{des07} found similar trends for
optically non-Seyfert ULIRGs, with the abscissa of 24 $\mu$m luminosity.  
} 
\end{figure}

\begin{figure}
\includegraphics[angle=-90,scale=.6]{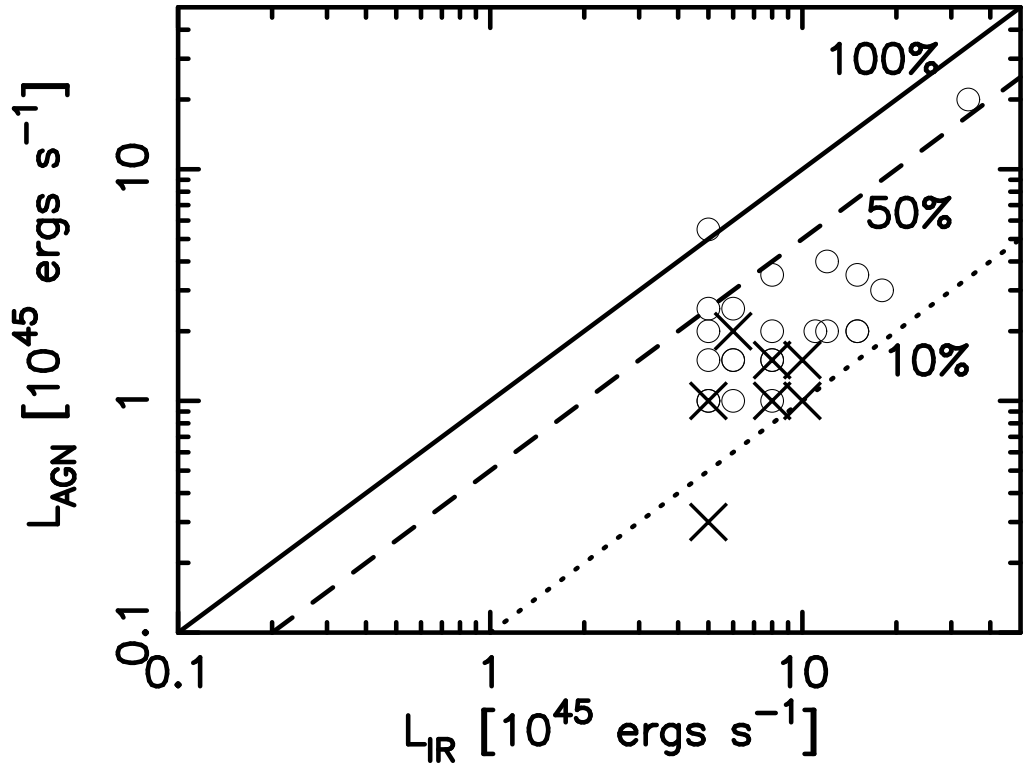} 
\caption{
Comparison of the intrinsic buried AGN luminosity and observed infrared
luminosity for selected ULIRGs with strong buried AGN signatures in their
spectra (i.e., PAH equivalent widths are particularly low and the
intrinsic AGN luminosities can be estimated with small ambiguities).
Optically non-Seyfert ULIRGs that were newly studied in this paper are plotted
with ``X'' symbols.  
Open circles represent sources studied in previously published papers 
\citep{ima07a,ima09}.
The solid line indicates that the buried AGN luminosity equals the 
infrared luminosity. In other words, the observed infrared luminosity can be
fully accounted for with a buried AGN.  
The dashed and dotted lines respectively indicate that 50\% and 10\% 
of the infrared luminosity can be explained by a buried AGN.} 
\end{figure}

\begin{figure}
\includegraphics[angle=-90,scale=.6]{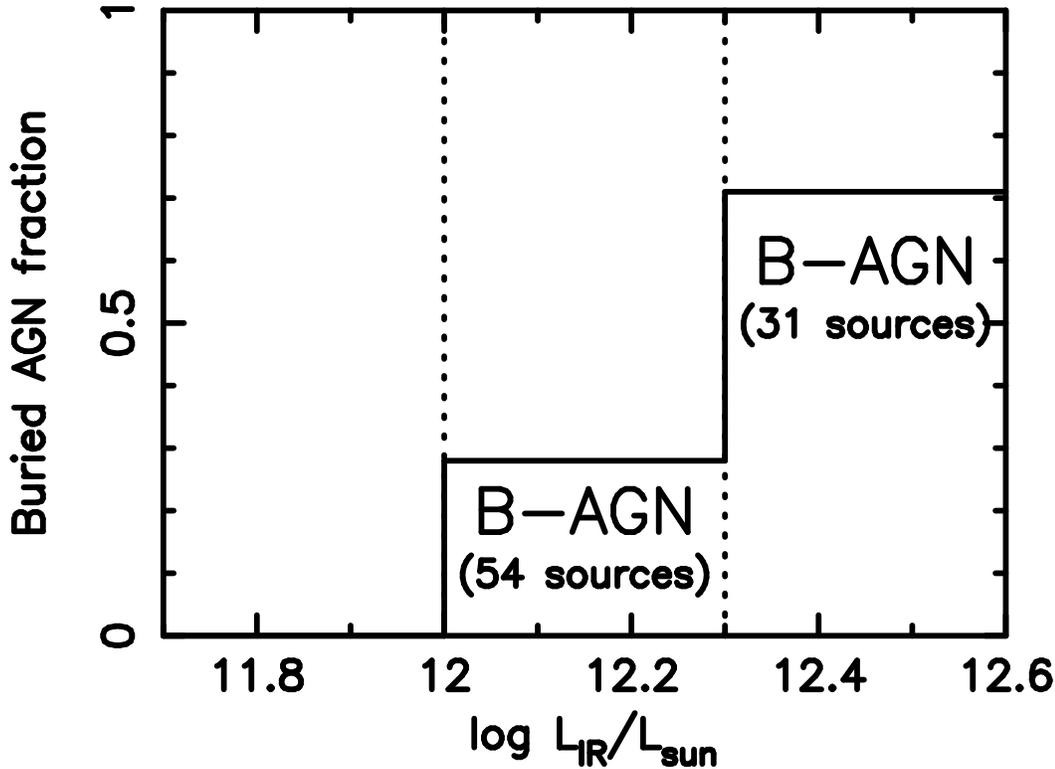} 
\caption{
Fraction of sources with strong buried AGN signatures as a
function of galaxy infrared luminosity. 
For ULIRGs (L$_{\rm IR}$ $\geq$ 10$^{12}$L$_{\odot}$), the fraction is
derived from the combination of \citet{ima07a}, \citet{ima09} and this
paper.  
The total number of sources is 54 for ULIRGs with 
10$^{12}$L$_{\odot}$ $\leq$ L$_{\rm IR}$ $<$ 10$^{12.3}$L$_{\odot}$, 
and 31 for those with L$_{\rm IR}$ $\geq$10$^{12.3}$L$_{\odot}$.  
For galaxies with L$_{\rm IR}$ $<$ 10$^{12}$L$_{\odot}$, the fraction is
obtained from the limited sample of 18 galaxies studied in \citet{bra06}, and
does not necessarily imply that no buried AGNs are present.
Although the sample is not statistically complete at 
L$_{\rm IR}$ $<$ 10$^{12}$L$_{\odot}$, there should not be a strong
bias, because the sources in this plot are essentially selected in the
same way, based on the {\it IRAS} 60 $\mu$m fluxes.} 
\end{figure}

\begin{figure}
\includegraphics[angle=0,scale=.6]{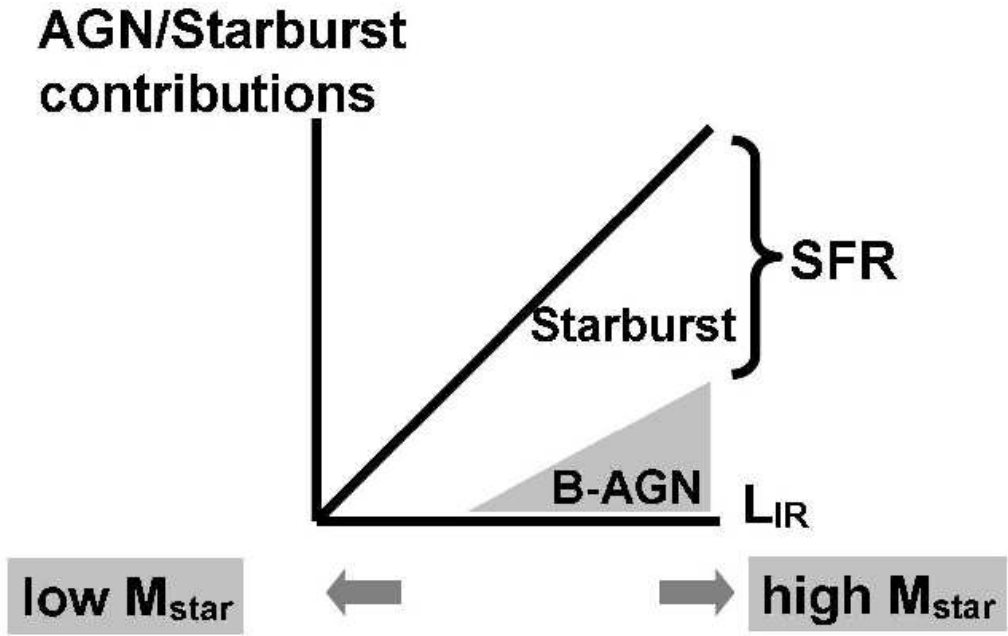} 
\caption{
Schematic diagram of the energetic importance of buried AGNs (B-AGN) and
starbursts as a function of galaxy infrared luminosity.
Both the abscissa and ordinate are plotted in a linear scale. 
SFR stands for a star formation rate.
In galaxies with higher infrared luminosities, 
the energetic importance of buried AGNs is {\it relatively} higher, and 
higher SFRs suggest that these galaxies will evolve into more massive
galaxies with larger stellar masses.} 
\end{figure}

\end{document}